\documentclass[final]{IEEEtran}

\usepackage{amsmath,amsfonts,amssymb,amsopn}
\usepackage{graphicx}
\usepackage{xcolor}
\usepackage{cite}
\usepackage{xspace}
\usepackage{subfig}
\usepackage{mathtools}
\usepackage{url}
\usepackage{algorithm}
\usepackage{algpseudocode}
\usepackage{ifpdf}

\ifpdf
  \usepackage[pdftex,colorlinks=true,breaklinks=true,citecolor=blue]{hyperref}
\else
  \usepackage[ps2pdf,colorlinks=true,citecolor=blue]{hyperref}
\fi

\usepackage{float}
\floatname{algorithm}{Procedure}

\ifCLASSOPTIONonecolumn

    \usepackage[margin=25mm]{geometry}
\else
    
\fi

\newcommand{\bv}[1]{\boldsymbol{#1}}
\newcommand{\B}[1]{\mathbf{#1}}
\newcommand{\dfn}{\triangleq}
\newcommand{\unit}[1]{\bv{\hat{#1}}}
\newcommand{\untsph}{\mathbb{S}^{2}}
\newcommand{\rotop}{\mathcal{D}(\varphi,\vartheta,\omega)}

\newcommand{\Dp}{\mathcal{D}_{\rho}}

\newcommand{\lsph}{L^2(\untsph)}

\newcommand{\lsphL}[1]{\mathcal{H}_{#1}}

\newcommand{\intsph}{\int_{\untsph}}
\newcommand{\inttheta}{\int_{\theta=0}^{\pi}}
\newcommand{\intphi}{\int_{\phi=0}^{2\pi}}
\newcommand{\SO}{\mathbb{SO}(3)}
\newcommand{\lSO}{L^{2}(\mathbb{SO}(3))}
\newcommand{\intSO}{\int_{\SO}}

\DeclarePairedDelimiterX{\norm}[1]{\lVert}{\rVert}{#1}

\newcommand{\innerp}[2]{\left \langle {#1} , {#2} \right \rangle}

\newcommand{\innerpS}[2]{\left \langle {#1} , {#2} \right \rangle_{\untsph}}
\newcommand{\innerpSO}[2]{\left \langle {#1} , {#2} \right \rangle_{\SO}}

\newcommand{\D}{\mathcal{D}}

\newcommand{\E}{\mathbb{E}}

\newcommand{\pr}{\Theta_c}

\newcommand{\R}{\mathbb{R}}
\newcommand{\N}{\mathbb{N}}

\newcommand{\matlab}{\texttt{MATLAB}}

\newcommand{\figref}[1]{Fig.\,\ref{#1}}

\newcommand{\secref}[1]{Section\,\ref{#1}}

\newcommand{\LSlp}{L_g}
\newcommand{\Lf}{L_f}
\newcommand{\sv}{\Sigma^2}

\newcommand{\bvr}{\mathrm {BVR}}

\newcommand{\res}{s}
\newcommand{\wav}[1]{\Psi^{(#1)}}

\newcommand{\wavcoef}[2]{w_{#1}^{\wav{#2}}}

\newtheorem{remark}{Remark}

\graphicspath{{figs/},{pdfs/},{pdf/},{Figures/},{Authors/}}

\begin{document}

\title{Spatial-Slepian Transform on the Sphere}

\author{%
Adeem~Aslam,~\IEEEmembership{Student Member,~IEEE}, and Zubair~Khalid,~\IEEEmembership{Senior Member,~IEEE}
 \thanks{A.~Aslam and Z.~Khalid are with the Department of Electrical Engineering, Syed Babar Ali School of Science and Engineering, Lahore University of Management Sciences, Lahore, Pakistan.}

\thanks{E-mail: adeem.aslam@lums.edu.pk, zubair.khalid@lums.edu.pk}
}

\maketitle

\begin{abstract}
We present spatial-Slepian transform~(SST) for the representation of signals on the sphere to support localized signal analysis. We use well-optimally concentrated Slepian functions, obtained by solving the Slepian spatial-spectral concentration problem of finding bandlimited and spatially optimally concentrated functions on the sphere, to formulate the proposed transform and obtain the joint spatial-Slepian domain representation of the signal. Due to the optimal energy concentration of the Slepian functions in the spatial domain, the proposed spatial-Slepian transform allows us to probe spatially localized content of the signal. Furthermore, we present an inverse transform to recover the signal from the spatial-Slepian coefficients, and show that well-optimally concentrated rotated Slepian functions form a tight frame on the sphere. We develop an algorithm for the fast computation of the spatial-Slepian transform and carry out computational complexity analysis. We present the formulation of SST for zonal Slepian functions, which are spatially optimally concentrated in the polar cap~(axisymmetric) region, and provide an illustration using the Earth topography map. To demonstrate the utility of the proposed transform, we carry out localized variation analysis; employing SST for detecting hidden localized variations in the signal.
\end{abstract}


\section{Introduction}
\label{sec:intro}

Spherical signal processing is the study and analysis of spherical signals, i.e., signals defined on the sphere, which are naturally encountered in many areas of science and engineering such as computer graphics~\cite{nadeem2016spherical}, medical imaging~\cite{michailovich:2009,Bates:2016,bergamasco20183d}, acoustics~\cite{Bates:2015,Liu:2019}, planetary sciences~\cite{Hoogenboom:2004,Audet:2014,Khaki:2018,Galanti:2019,Hippel:2019}, geophysics~\cite{Wieczorek:2005,Simons:2006Polar}, cosmology~\cite{Dahlen:2008,Marinucci:2008,McEwen:2012}, quantum mechanics~\cite{grinter2018}, wireless communications~\cite{Kennedy:2013-ICSPCS, Alem:2015, bashar:2016} and antenna design~\cite{talashila:2019}, to name a few. A natural choice of basis functions for the representation of signals on the sphere are the spherical harmonic functions~(or spherical harmonics for short). Such a representation is enabled by the spherical harmonic transform~(SHT) and is called the spherical harmonic~(or spectral) domain representation.

The representation of a signal in the spectral domain reveals global characteristics of the signal, without any regards to the scale or localization of those characteristics. In order to probe signals at different scales, more sophisticated methods have been proposed in the literature. One such tool that has been extensively used to represent time domain signals at different scales is the wavelet transform~\cite{Mallat:1989,Daubechies:1990,Mallat-book:2009}, which has also been extended for signal analysis on the sphere~\cite{Narcowich:1996,Freeden:1997,Antoine:1999,Starck:2006,Wiaux:2008,McEwen:2018}. The framework of wavelet transform uses wavelet functions to record scale-dependent information of the underlying signal in what are called as wavelet coefficients. Although the wavelet functions have been shown to exhibit good spatial localization~\cite{McEwen:2018}, they cannot be adapted to the shape of the region of interest on the sphere. Consequently, for applications where signal is to be analyzed locally over a region on the sphere, it is imperative to find alternate methods which can be used to probe local characteristics of signals over a subset of the sphere.

Motivated by the idea of wavelet transform, where the signal content is essentially spread out in the joint space-scale domain, we seek to find a representation of signals to analyze their local characteristics in an effort to detect localized hidden features. Naturally, we revert to the Slepian spatial-spectral concentration problem on the sphere~\cite{Albertella:1999,Simons:2006,Bates2:2017}, which results in optimally localized basis functions, called Slepian functions, that can be used for accurate representation and reconstruction of the underlying signal in a given region on the sphere. Using well-optimally concentrated Slepian functions, with varying energy concentration with in a region on the sphere, we propose a transform, referred to as spatial-Slepian transform, which is similar in spirit to the wavelet transform but uses bandlimited and spatially well-optimally concentrated Slepian functions instead of wavelet functions. Unlike the wavelet transform, spatial-Slepian transform probes local content of the signal, which is a direct consequence of the use of well-optimally concentrated Slepian functions. The number of resulting spatial-Slepian coefficients is determined by the fractional area of the region on the sphere, which is chosen to solve the spatial-spectral concentration problem. In this context, the main contribution of this work is summarized below:
\begin{itemize}
	\item We use bandlimited and spatially well-optimally concentrated Slepian functions to formulate the proposed spatial-Slepian transform~(SST) as the inner product between the signal and the rotated Slepian functions in \secref{sec:SST}, where we also present the inverse transform to recover the signal from its spatial-Slepian coefficients and show that the well-optimally concentrated rotated Slepian functions form a tight frame for the Hilbert space of bandlimited functions on the sphere. Furthermore, we present analytical expressions for the spatial-Slepian coefficients, computed over axisymmetric north polar cap region using zonal Slepian functions, and present an illustration on the Earth topography map.
	\item We develop an algorithm for the fast computation of SST in \secref{sec:SST}, analyze the computational complexity of the algorithm and validate the results for a test signal which is synthesized in the spherical harmonic domain.
	\item In \secref{sec:LVA}, we present an application of the proposed SST by developing a framework for the detection of hidden localized variations in the signal. We compare the results obtained using the proposed transform with those obtained from the wavelet transform and show that spatial-Slepian transform performs better by achieving a better estimate of the underlying region of the hidden localized variations.
\end{itemize}

Before presenting the proposed work, we review the necessary mathematical background for signal analysis on the sphere and briefly discuss the spatial-spectral concentration problem in the next section.

\section{Mathematical Background}
\label{sec:maths}

\subsection{Signals on $2$-Sphere}
We consider complex valued and square-integrable functions on the surface of the $2$-sphere~(sphere for short) which is defined as $\untsph \dfn
\{\unit{x} \in \mathbb{R}^3 \colon |\unit{x}| = 1  \}$, where $|\cdot|$ denotes the Euclidean norm, $\unit{x} \equiv \unit{x}(\theta,\phi) \dfn (\sin\theta\,\cos\phi,\, \sin\theta\,\sin\phi, \, \cos\theta)^{\mathrm{T}}$ is the unit vector in $\mathbb{R}^3$, parameterized by the colatitude angle, $\theta \in [0, \pi]$ measured from the positive $z$-axis, and longitude angle, $\phi\in [0, 2\pi)$ measured from the positive $x$-axis in the $x-y$ plane, and $(\cdot)^{\mathrm{T}}$ denotes the transpose operation. We denote such functions by $f(\unit{x}) \equiv f(\theta, \phi)$ and define the inner product between any two functions $f, h$ as~\cite{Kennedy-book:2013}
\begin{align}
\innerpS{f}{h} \dfn \int_{\untsph}
f(\unit{x}) \overline{h(\unit{x})} \, ds(\unit{x}),
\label{eq:inner_prod}
\end{align}
where $\overline{(\cdot)}$ denotes the complex conjugate, $ds(\unit{x}) \equiv \sin\theta\,d\theta\,d\phi$ is the differential area element on the sphere and the integration is carried out over the whole sphere, i.e., $\displaystyle \int_{\untsph} = \inttheta\intphi$. Equipped with the inner product in \eqref{eq:inner_prod}, set of complex-valued, square-integrable functions on the sphere forms a Hilbert space, denoted by $\lsph$. Norm of the function $f$ is induced by the inner product as $\norm{f}_{\untsph} \dfn \innerpS{f}{f}^{1/2}$ and its energy is given by $\norm{f}^2_{\untsph}$. Functions with finite energy are referred to as signals on the sphere. For a given spatial region $R \subset \untsph$, we also define
\begin{align}
\innerp{f}{h}_R = \int_R f(\unit{x}) \overline{h(\unit{x})} \, ds(\unit{x})
\label{eq:inner_prod_R}
\end{align}
as the local inner product between $f$ and $h$, where $\|f\|^2_{R} \dfn \innerp{f}{f}$ quantifies the energy of signal $f$ in the region $R$.

The Hilbert space $\lsph$ is separable and contains a complete set of orthonormal basis functions called spherical harmonics, given by~\cite{Kennedy-book:2013}
\begin{align*}
Y_{\ell}^m(\unit{x}) \equiv Y_{\ell}^m(\theta, \phi) \dfn \sqrt{\frac{2\ell + 1}{4\pi} \frac{(\ell-m)!}{(\ell+m)!}} P_{\ell}^m(\cos\theta)e^{im\phi}
\end{align*}
for integer degree $\ell \geq 0$ and integer order $|m| \leq \ell$, where $P_{\ell}^m(\cos\theta)$ is the associated Legendre polynomial of degree $\ell$ and order $m$~\cite{Kennedy-book:2013}. As a result, any signal $f \in \lsph$ can be expanded as
\begin{align}
f(\theta,\phi) = \sum_{\ell,m}^{\infty} (f)_{\ell}^m Y_{\ell}^m(\theta,\phi),
\label{eq:f_expansion}
\end{align}
where we have used the shorthand notation $\sum\limits_{\ell,m}^{\infty} \equiv \sum\limits_{\ell=0}^{\infty} \sum\limits_{m=-\ell}^{\ell}$, and
\begin{align}
(f)^{\ell}_m \dfn \innerpS{f}{Y_{\ell}^m} = \intsph f(\theta,\phi) \overline{Y_{\ell}^m(\theta,\phi)} \, \sin\theta d\theta d\phi
\label{eq:flm}
\end{align}
is the spherical harmonic~(spectral) coefficient of degree $\ell$ and order $m$, which forms the spherical harmonic~(spectral) domain representation of the signal $f$. Signal $f \in \lsph$ is called bandlimited to degree $\Lf$ if $(f)_{\ell}^m = 0$ for $\ell, |m| \geq \Lf$. Set of all such bandlimited signals on the sphere forms an $L^2$-dimensional subspace of $\lsph$, denoted by $\lsphL{\Lf}$, and their spectral coefficients can be stored in an $\Lf^2 \times 1$ column vector as
\begin{align}
\B{f} = \left[(f)_0^0,(f)_1^{-1},(f)_1^0,(f)_1^1,\ldots,(f)_{\Lf-1}^{\Lf-1}\right]^{\mathrm{T}}.
\label{eq:flm_column}
\end{align}

\subsection{Signal Rotation on the Sphere}
A point on the surface of the sphere can be rotated to any given orientation by sequential application of sub-rotations by $\omega \in [0,2\pi)$ around $z$-axis, $\vartheta \in [0,\pi]$ around $y$-axis and $\varphi \in [0,2\pi)$ around $z$-axis, following right-handed convention. The angles $\omega$, $\vartheta$ and $\varphi$ are called Euler angles. Each rotation by an Euler angle is represented by a $3 \times 3$ orthogonal rotation matrix and the overall rotation is specified by a matrix $\B{R}$ defined as
\begin{align}
\B{R} \equiv \B{R}(\varphi,\vartheta,\omega) \dfn \B{R}_{z}(\varphi) \B{R}_{y}(\vartheta) \B{R}_{z}(\omega)
\label{eq:rot_matrix}
\end{align}
where $\B{R}_{z}(\omega)$ and $\B{R}_{y}(\vartheta)$ are the matrices representing rotations by angles $\vartheta$ around $y$-axis and $\omega$ around $z$-axes respectively~\cite{Kennedy-book:2013}.

Defining $\rho$ as the $3$-tuple of Euler angles, i.e., $\rho \dfn (\varphi, \vartheta, \omega)$, signal rotation on the sphere is specified by a rotation operator $\Dp \equiv \rotop$, whose action on a signal $f \in \lsph$ is defined as the inverse rotation of the coordinate system, i.e.,
\begin{align}
(\Dp f)(\unit{x}) \equiv (\rotop f)(\unit{x}) \dfn f(\B{R}^{-1}\unit{x}),
\label{eq:rot_action}
\end{align}
where $\B{R}$ is the rotation matrix in \eqref{eq:rot_matrix}. Spectral coefficients of the rotated signal are given by~\cite{Kennedy-book:2013}
\begin{align}
\left(\Dp f\right)_{\ell}^m = \sum\limits_{m' = -\ell}^{\ell} D^{\ell}_{m,m'}(\varphi, \vartheta, \omega) (f)_{\ell}^{m'},
\label{eq:rotated_flms}
\end{align}
where $D^{\ell}_{m,m'}(\varphi, \vartheta, \omega)$ is the Wigner-$D$ function defined as
\begin{align}
D^{\ell}_{m,m'}(\varphi, \vartheta, \omega) \dfn e^{-im\varphi} d_{m,m'}^{\ell}(\vartheta) e^{-im'\omega},
\label{eq:WignerD}
\end{align}
for degree $\ell$ and orders $|m|, |m'| \le \ell$, and $d_{m,m'}^{\ell}(\vartheta)$ is the Wigner-$d$ function~\cite{Kennedy-book:2013}. As a result, the rotated signal is given by
\begin{align}
\left(\Dp f\right)(\unit{x}) = \sum_{\ell,m,m'}^{\infty} D^{\ell}_{m,m'}(\varphi, \vartheta, \omega) (f)_{\ell}^{m'} Y_{\ell}^m(\theta,\phi).
\label{eq:rotated_f}
\end{align}

\subsection{Signals on the $\SO$ Rotation Group}
Group of all proper rotations\footnote{An improper rotation is a reflection or a flip about either some axes or the center of the coordinate system.}, represented by the $3$-tuple $\rho = (\varphi, \vartheta, \omega)$, is called the Special Orthogonal group, denoted by $\SO$. Square-integrable and complex-valued functions defined on the rotation group $\SO$ form a Hilbert space $\lSO$, such that the inner product between any two functions $v, w \in \lSO$ is given by
\begin{align}
\innerpSO{f}{h} \dfn \int_{\SO} v(\rho) \overline{w(\rho)} \, d\rho,
\label{eq:inner_prod_SO3}
\end{align}
where $d\rho \equiv d\varphi\sin\vartheta\, d\vartheta\, d\omega$ is the differential element on the $\SO$ rotation group and integration is carried out over all possible rotations, i.e., $\displaystyle \int_{\SO} = \int_{\varphi = 0}^{2\pi} \int_{\vartheta = 0}^{\pi} \int_{\omega = 0}^{2\pi}$. Inner product in \eqref{eq:inner_prod_SO3} induces a norm on the function $v \in \lSO$ as $\norm{v}_{\SO} \dfn \innerpSO{v}{v}^{1/2}$ and its energy is given by $\norm{v}^2_{\SO}$. Such finite energy functions are referred to as signals on the rotation group.

The Hilbert space $\lSO$ is separable and has Wigner-$D$ functions as the basis functions which admit the following orthogonality relation~\cite{Kennedy-book:2013, Sakurai:1994}
\begin{align}
\innerpSO{D^{\ell}_{m,m'}} {D^p_{q,q'}} = \left(\frac{8\pi^2}{2\ell+1}\right) \delta_{\ell,p} \delta_{m,q} \delta_{m',q'},
\label{eq:WignerD_ortho}
\end{align}
where $\delta_{m,n}$ is the Kronecker delta function. Therefore, any signal $v \in \lSO$ can be expanded as
\begin{align}
v(\rho) = \sum_{\ell,m,m'}^{\infty} (v)^{\ell}_{m,m'} D^{\ell}_{m,m'}(\rho),
\end{align}
where we have introduced the shorthand notation $\sum\limits_{\ell,m,m'}^{\infty} \equiv \sum\limits_{\ell=0}^{\infty}\sum\limits_{m=-\ell}^{\ell} \sum\limits_{m'=-\ell}^{\ell}$ and
\begin{align}
(v)^{\ell}_{m,m'} \dfn \left(\frac{2\ell+1}{8\pi^2}\right) \innerpSO{v}{D^{\ell}_{m,m'}}
\label{eq:flmm'}
\end{align}
is the $\SO$ spectral coefficient of degree $\ell$ and orders $m,m'$, constituting the spectral domain representation of the signal $v$. Signal $v$ is called bandlimited to degree $L_v$ if $(v)^{\ell}_{m,m'} = 0$ for all $\ell, |m|, |m'| \ge L_v$.

\subsection{Spatial-Spectral Concentration on the Sphere}
\label{sec:slep}

\begin{figure*}[!t]
	\centering
	\subfloat[$g_1(\unit{x})$]{
		\includegraphics[width=0.15\textwidth]{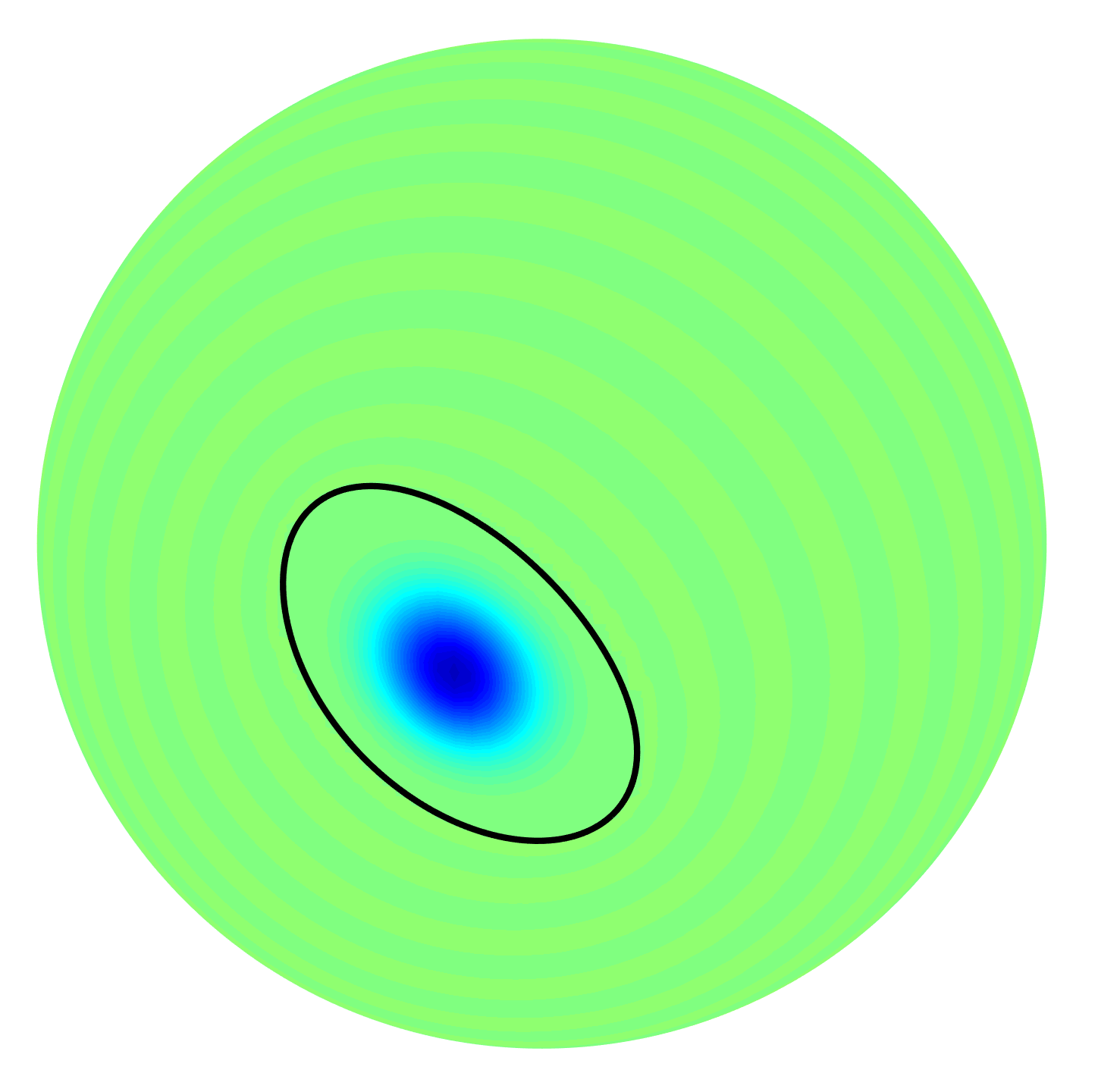}}\hfil
	\subfloat[$g_2(\unit{x})$]{
		\includegraphics[width=0.15\textwidth]{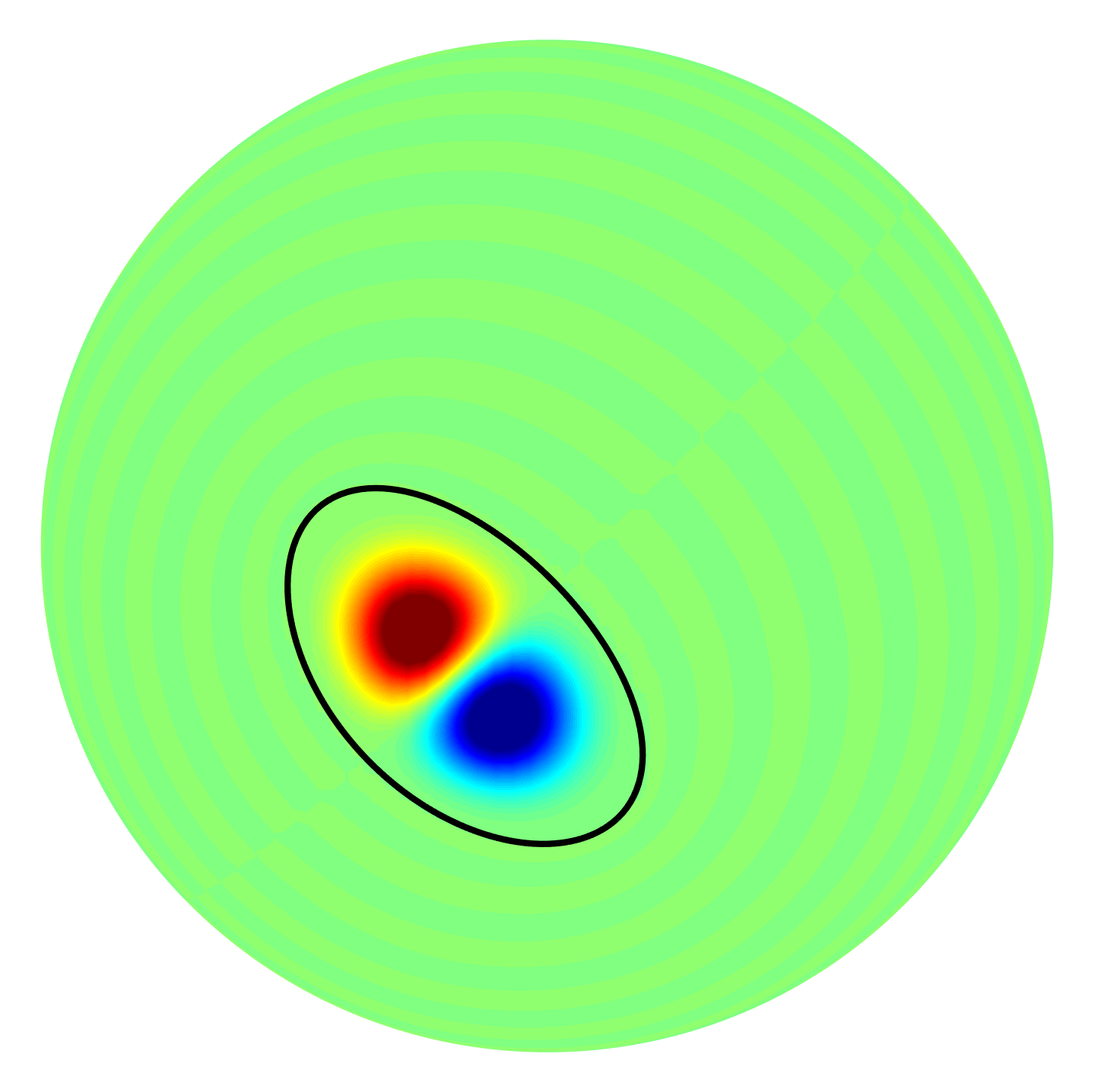}}\hfil
	\subfloat[$g_3(\unit{x})$]{
		\includegraphics[width=0.15\textwidth]{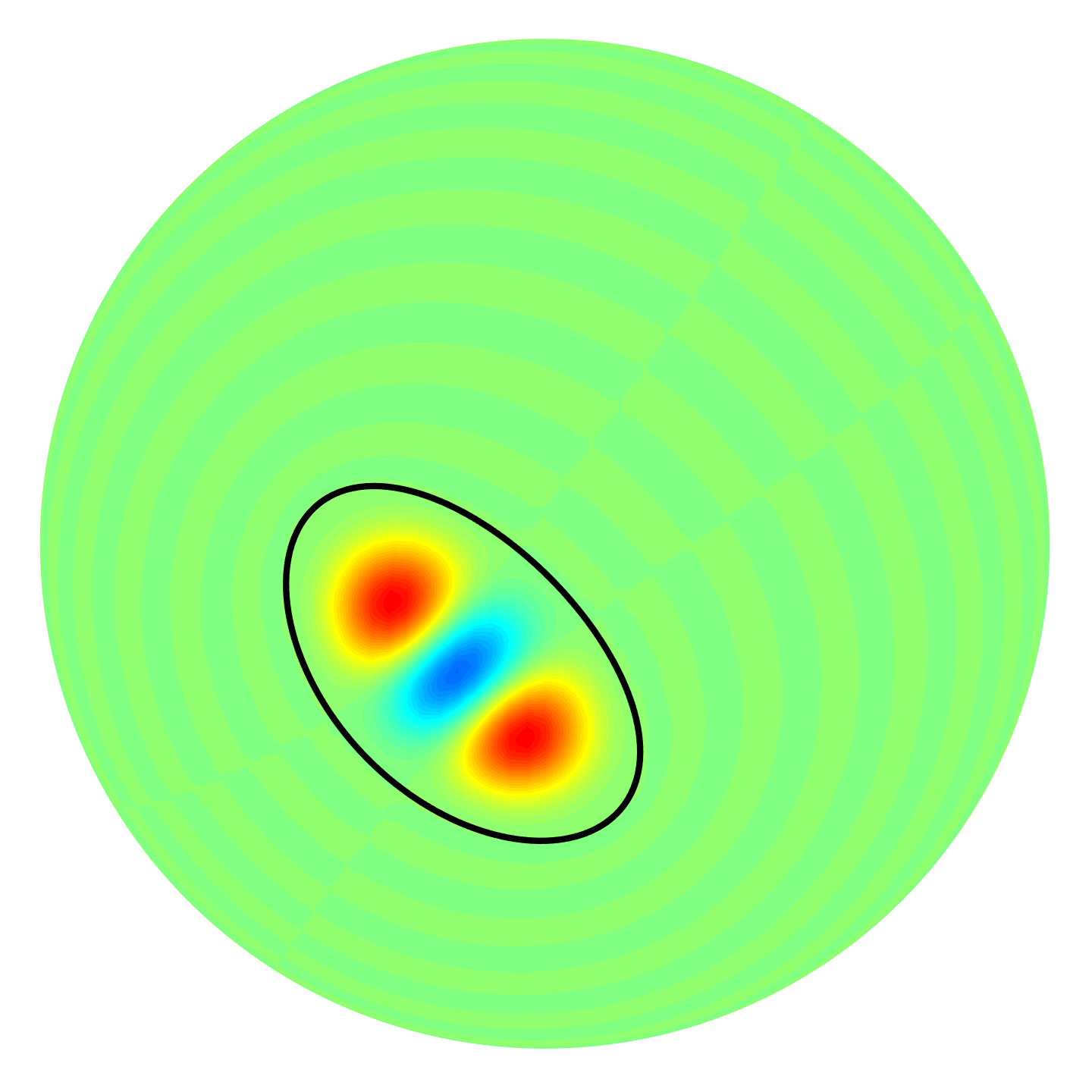}}\hfil
	\subfloat[$g_4(\unit{x})$]{
		\includegraphics[width=0.15\textwidth]{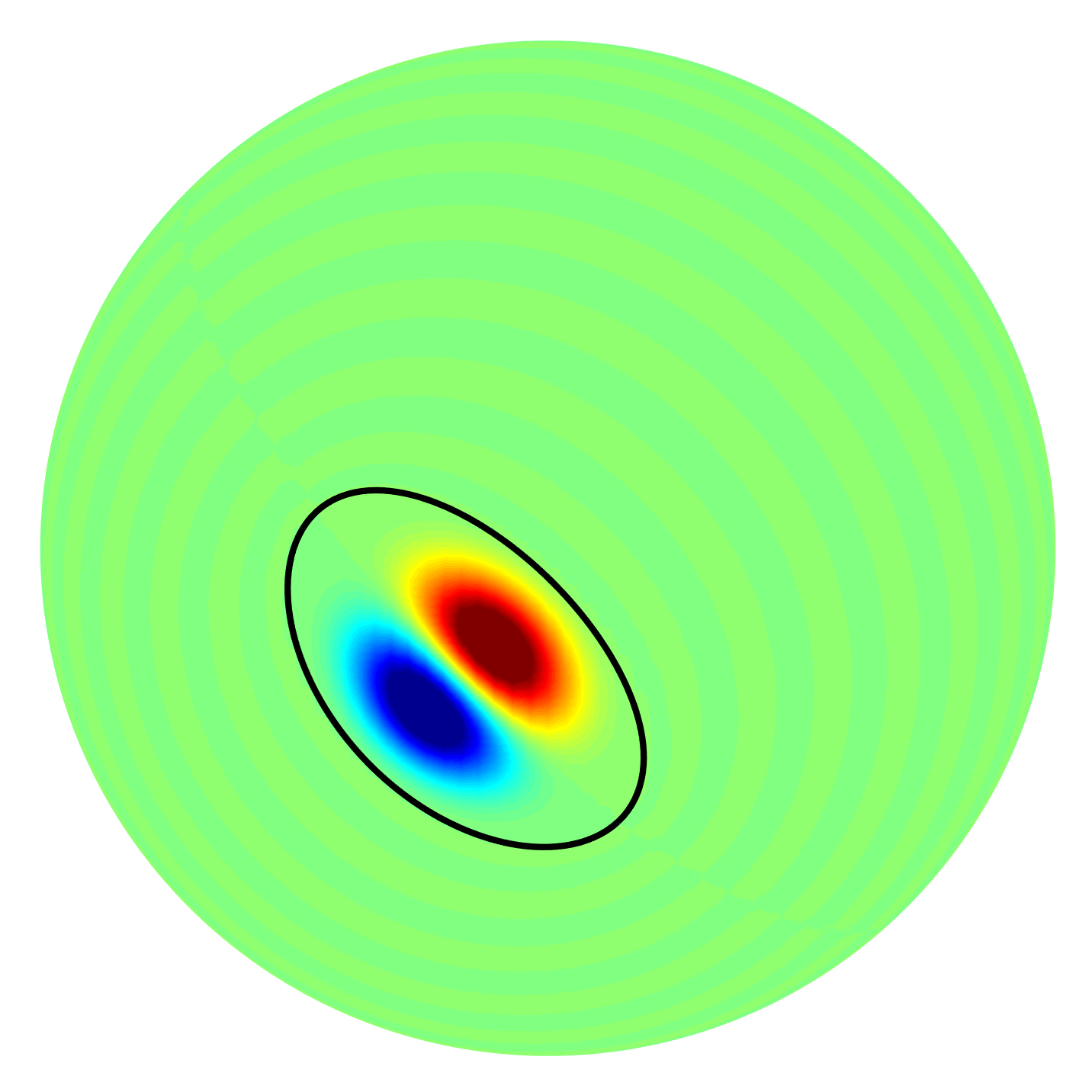}}\hfil	
	\subfloat[$g_5(\unit{x})$]{
		\includegraphics[width=0.15\textwidth]{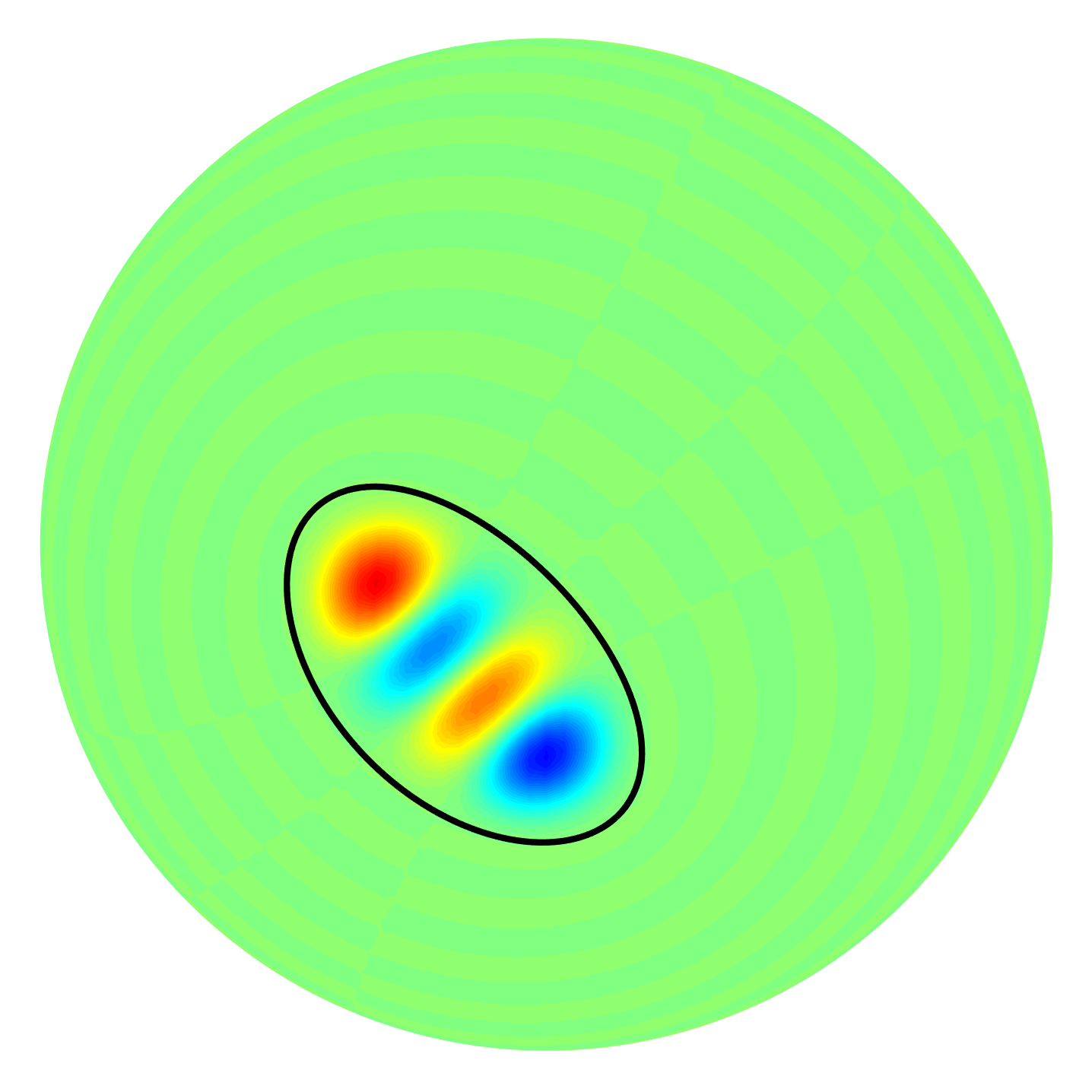}}\hfil	
	\subfloat[$g_6(\unit{x})$]{
		\includegraphics[width=0.15\textwidth]{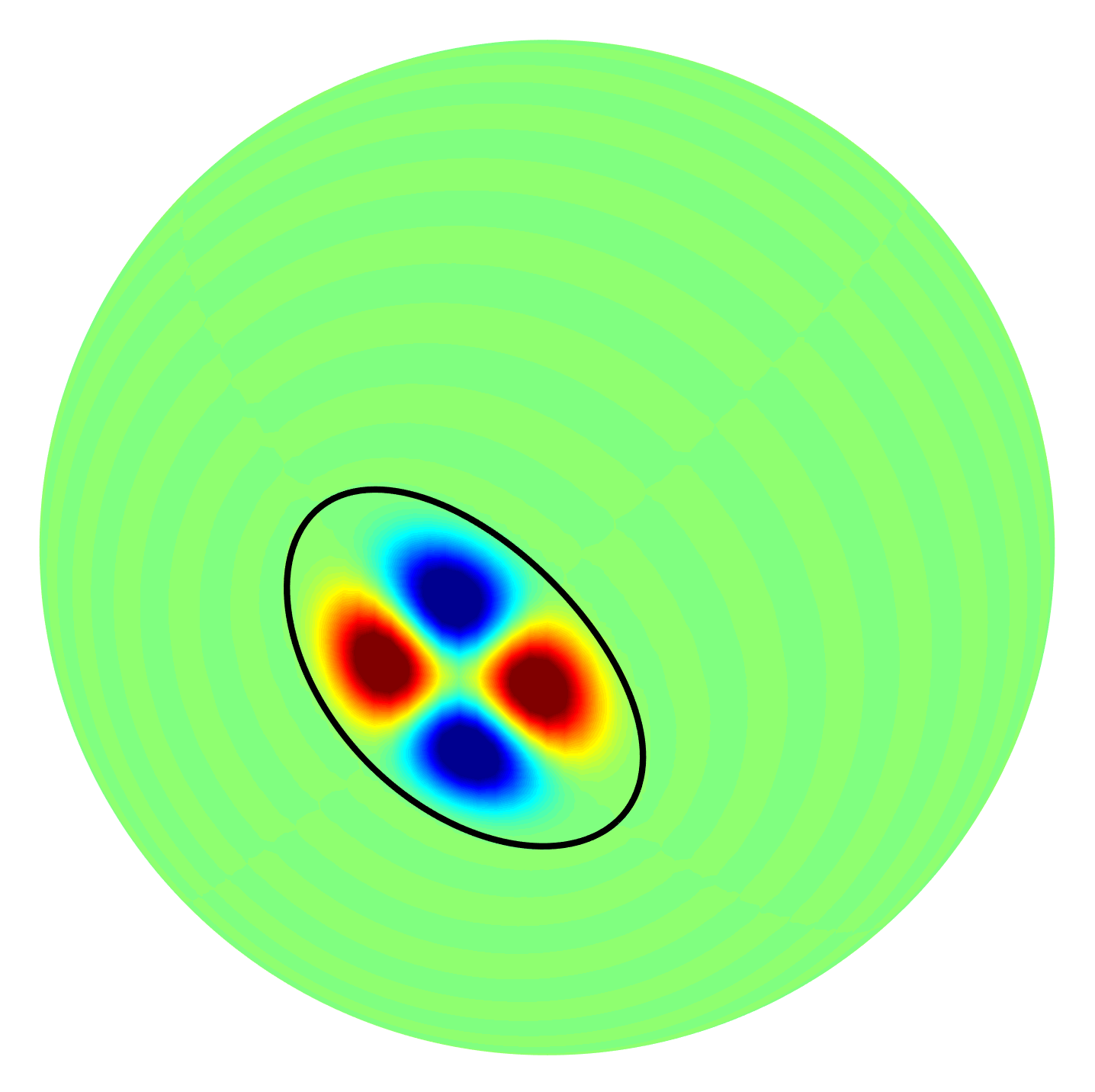}}\hfil
	
	\subfloat[$g_7(\unit{x})$]{
		\includegraphics[width=0.15\textwidth]{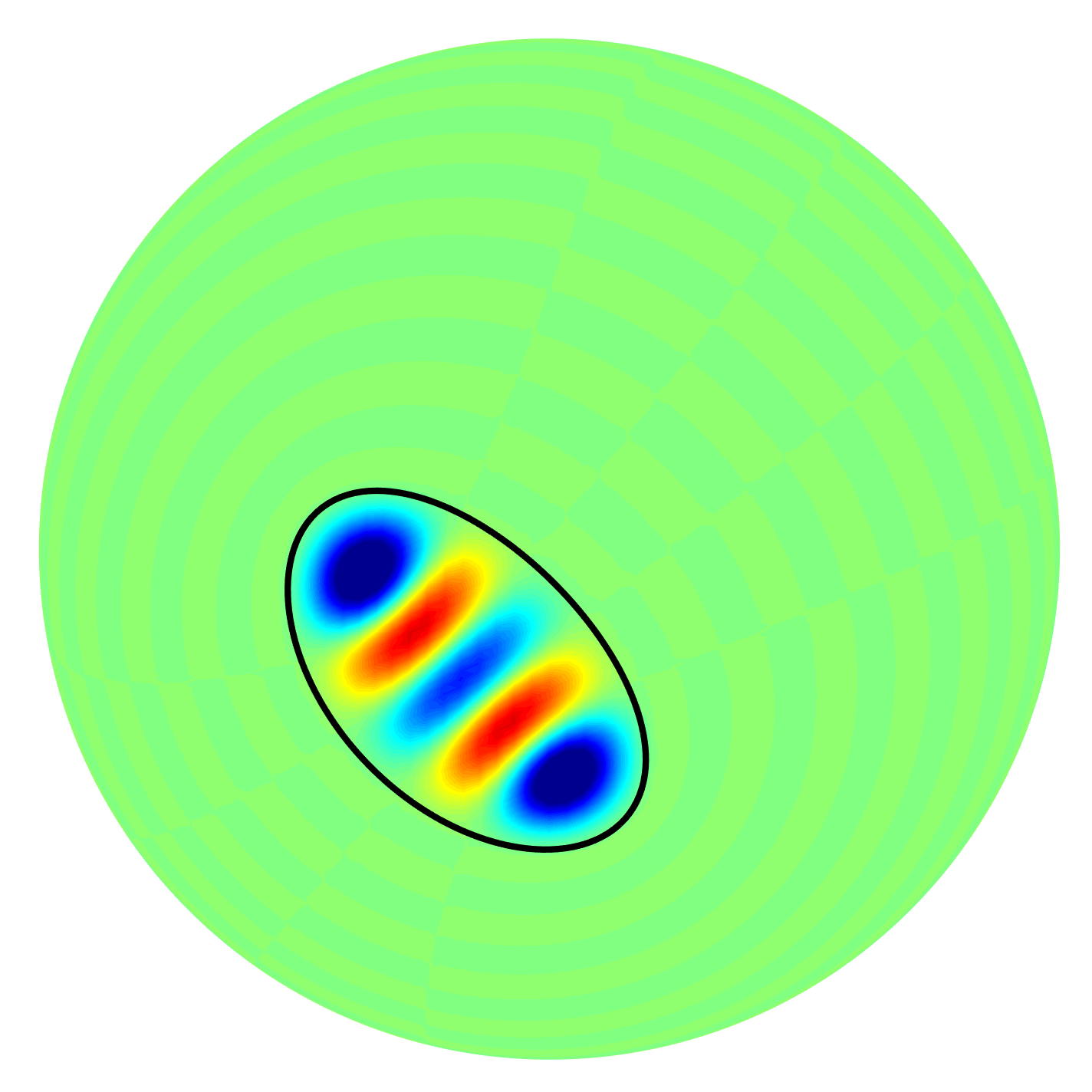}}\hfil	
	\subfloat[$g_8(\unit{x})$]{
		\includegraphics[width=0.15\textwidth]{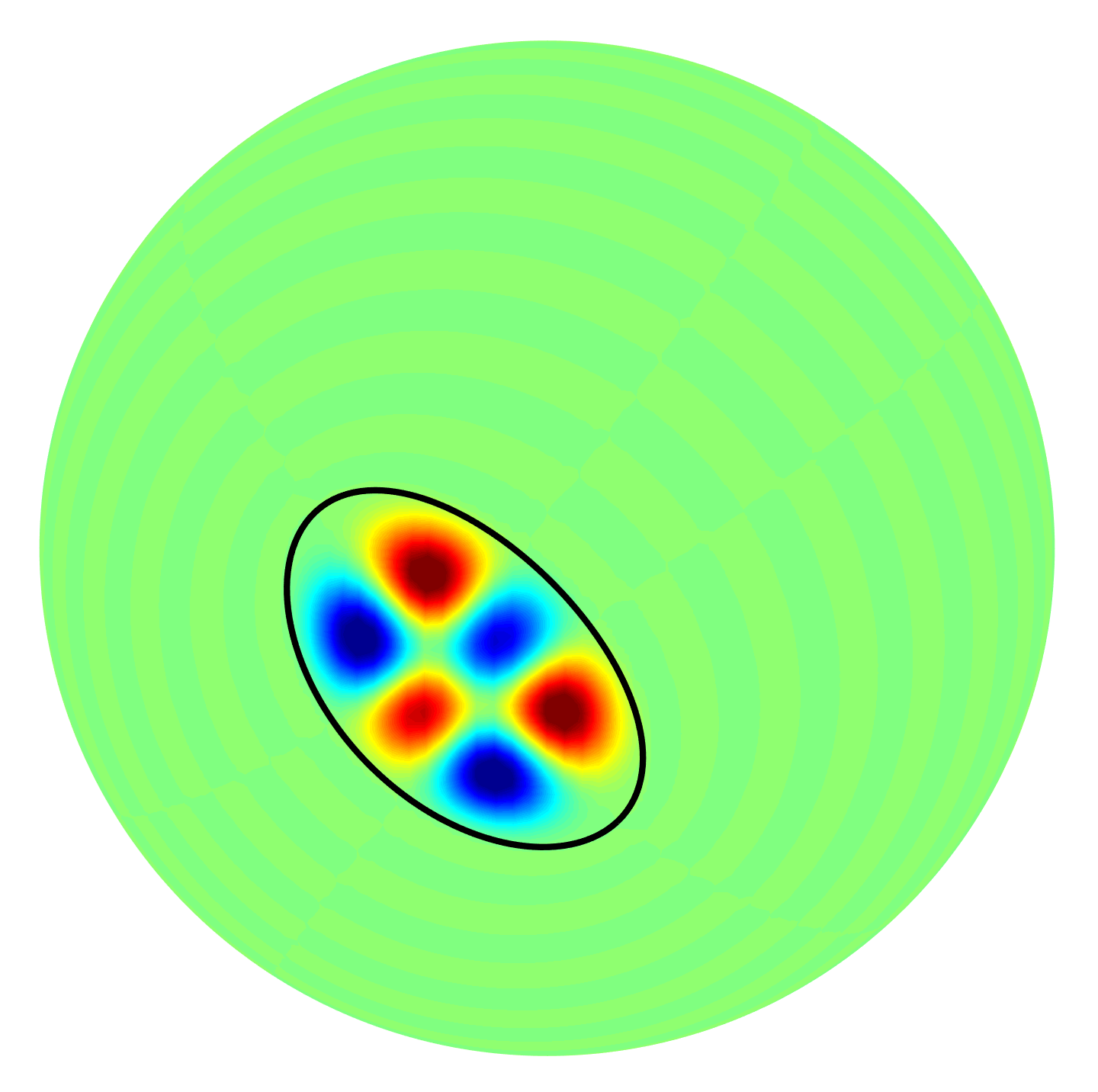}}\hfil
	\subfloat[$g_9(\unit{x})$]{
		\includegraphics[width=0.15\textwidth]{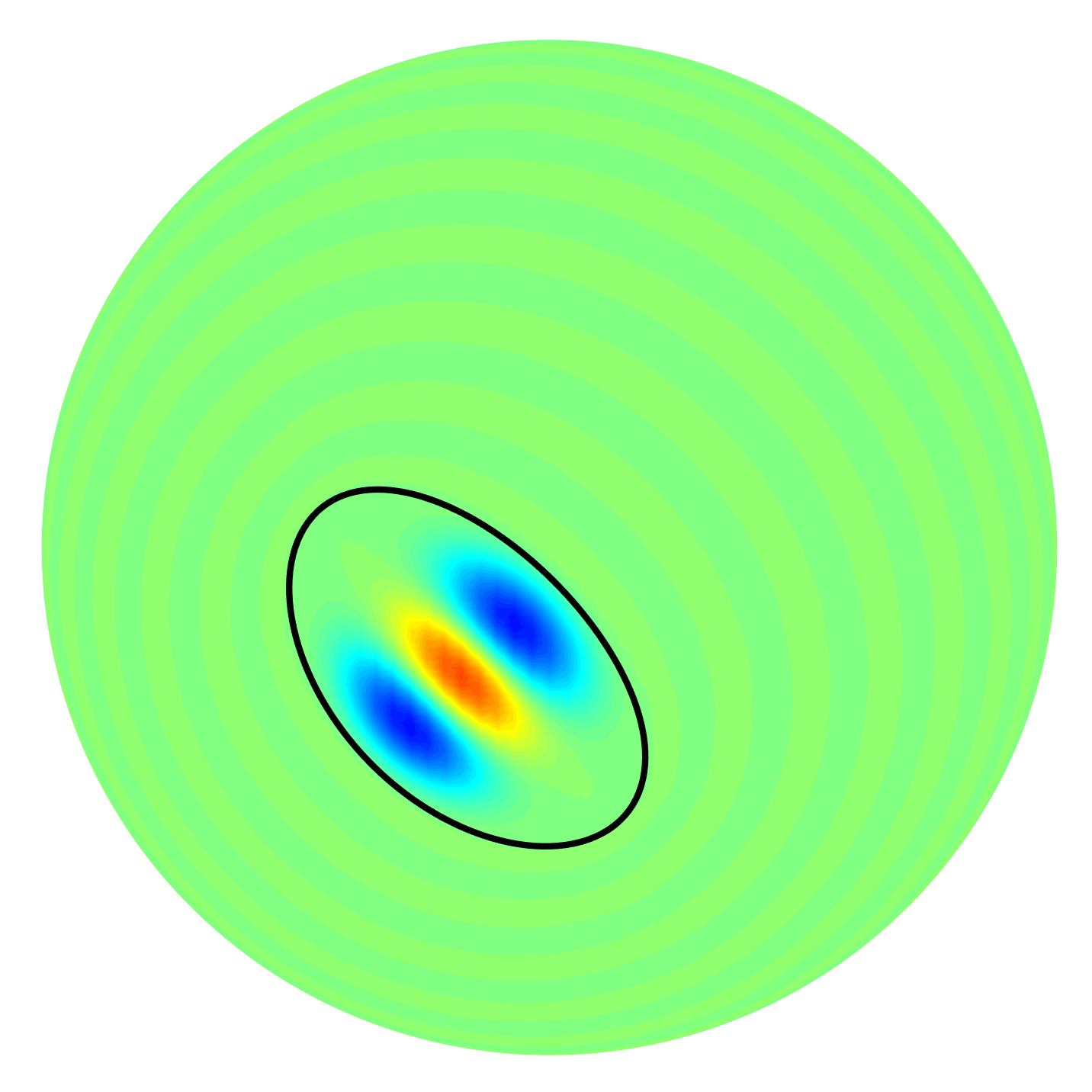}}\hfil
	\subfloat[$g_{10}(\unit{x})$]{
		\includegraphics[width=0.15\textwidth]{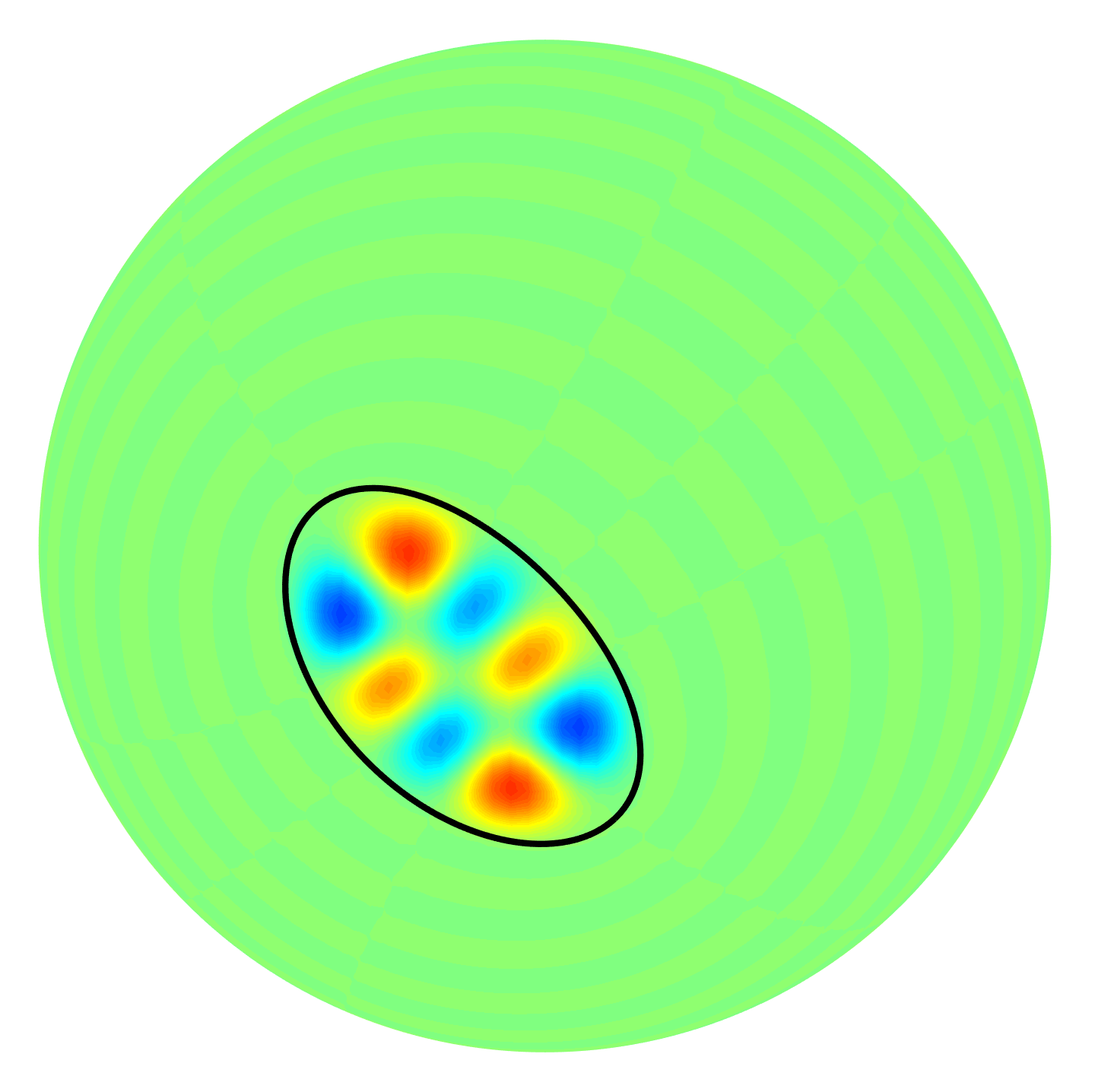}}\hfil
	\subfloat[$g_{11}(\unit{x})$]{
		\includegraphics[width=0.15\textwidth]{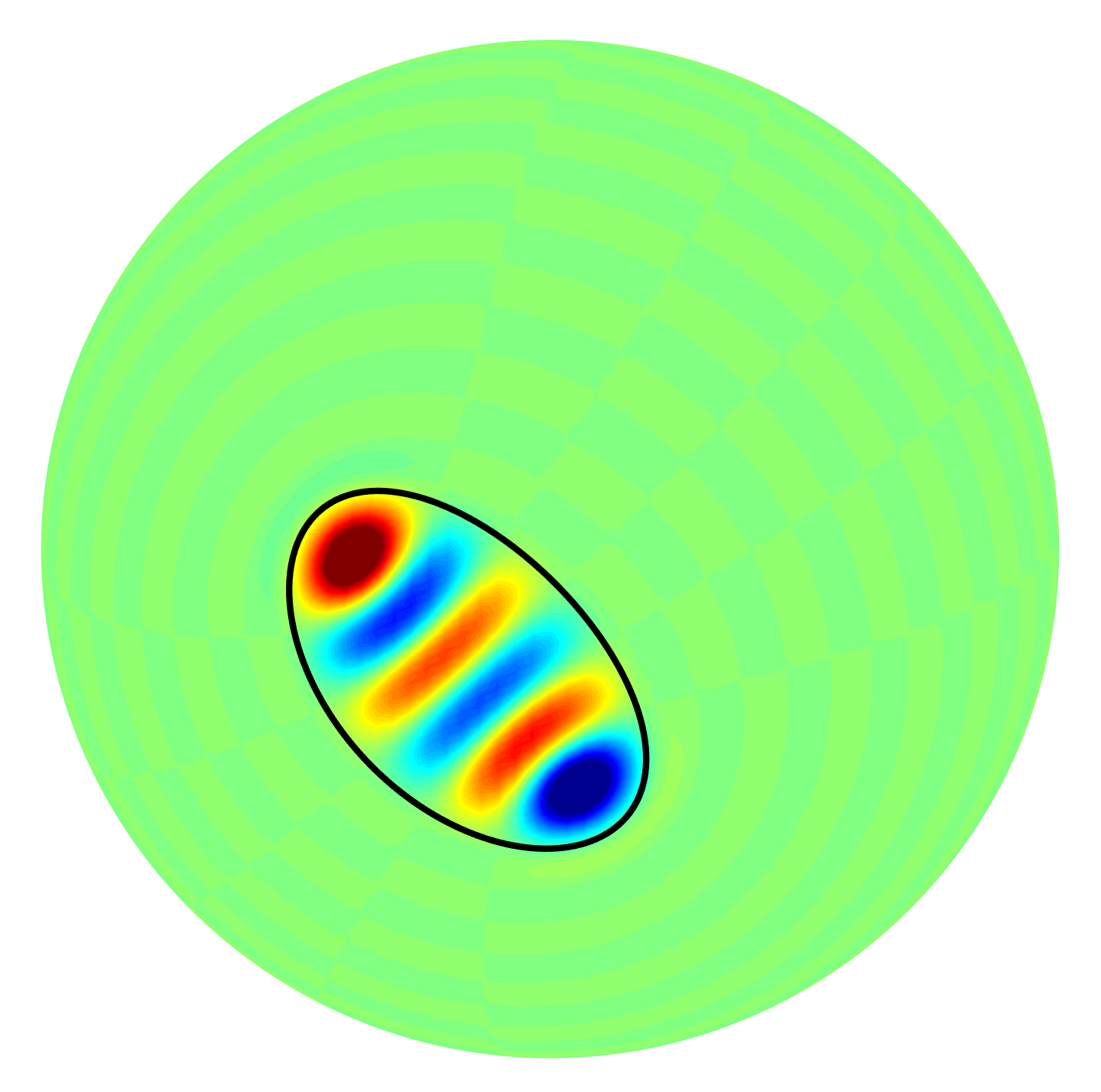}}\hfil
	\subfloat[$g_{12}(\unit{x})$]{
		\includegraphics[width=0.15\textwidth]{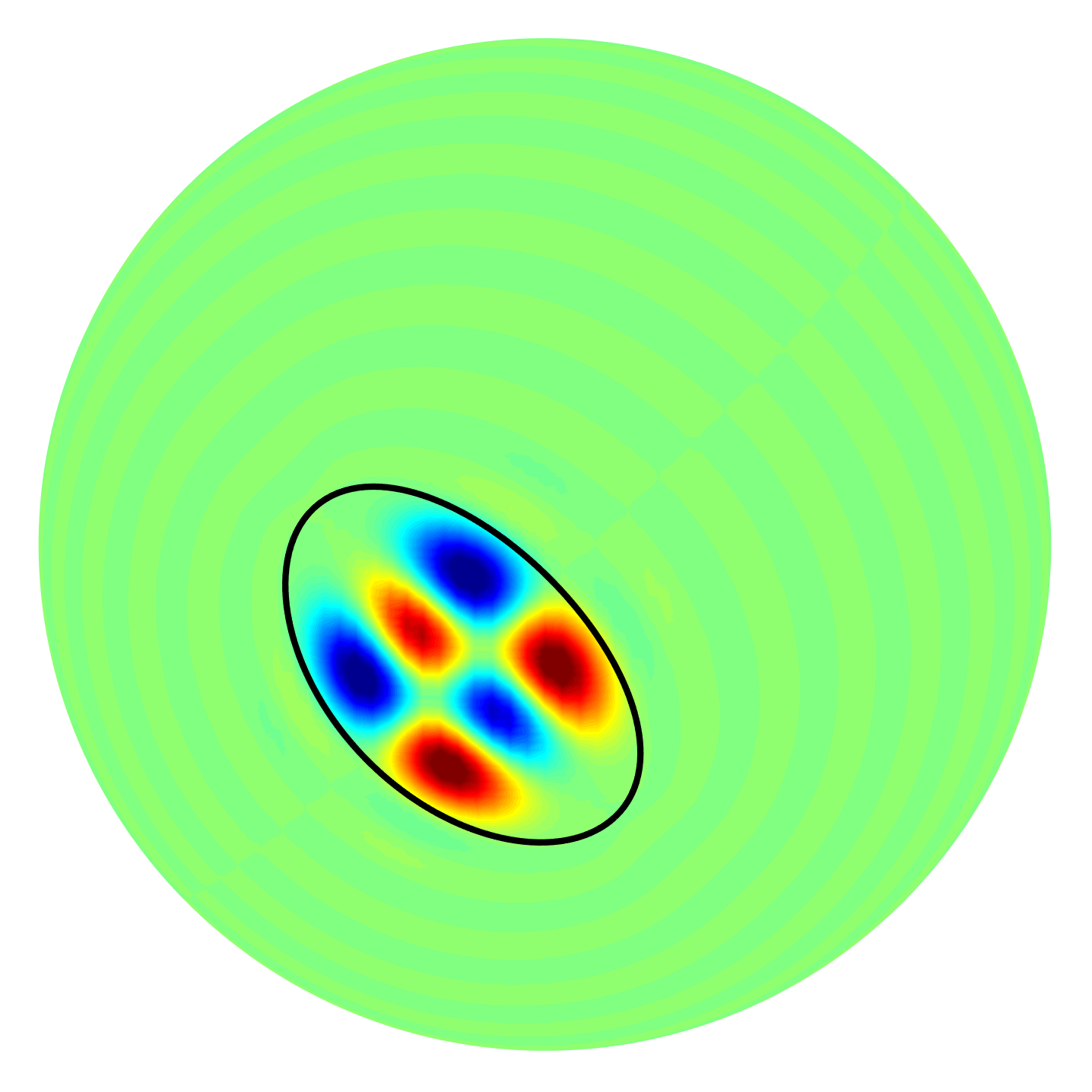}}\hfil
	
	\subfloat{
		\includegraphics[width=\textwidth]{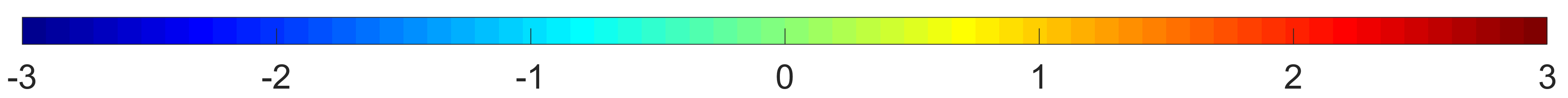}}\hfil
	\caption{Slepian functions computed over a rotated spherical ellipse which was initially aligned with $x$-axis, having focus colatitude $\theta_c = 15^{\circ}$ and semi-arc-length of the semi-major axis $a = 20^{\circ}$. The rotation angles are $\rho = (60^{\circ},90^{\circ},45^{\circ})$ and bandlimit $\LSlp = 32$.}
	\label{fig:Slepian_functions}
\end{figure*}

The problem of spatial concentration of bandlimited signals~(or equivalently spectral concentration of spatially limited signals) was first investigated by Slepian and his co-authors in their seminal work on time domain signals in $1960$s. They optimized a quadratic energy concentration measure to obtain an orthogonal family of strictly bandlimited signals which were optimally concentrated with in a given time interval~\cite{Slepian:1960}. This work was later extended to multidimensional Euclidean domain signals~\cite{Slepian:1964, Simons:2011} and for signals defined on the sphere~\cite{Albertella:1999, Simons:2006, Simons:2006Polar, Bates:2017, Bates2:2017}. In this section, we present a brief overview of the spatial concentration of bandlimited signals on the sphere.

To maximize the spatial energy concentration of a bandlimited signal $g \in \lsphL{\LSlp}$ in the spatial region $R\subset\untsph$, we optimize the following energy concentration
\begin{align}
\lambda &= \frac{\norm{g}^2_R}{\norm{g}^2_{\untsph}} = \frac{ \displaystyle \int\limits_R \sum\limits_{p,q}^{\LSlp-1} (g)_p^q Y_p^q(\unit{x}) \overline{ \left(\sum\limits_{\ell,m}^{\LSlp-1} (g)_{\ell}^m Y_{\ell}^m(\unit{x})\right) } }{ \displaystyle \int\limits_{\untsph} \sum\limits_{p,q}^{\LSlp-1} (g)_p^q Y_p^q(\unit{x}) \overline{ \left(\sum\limits_{\ell,m}^{\LSlp-1} (g)_{\ell}^m Y_{\ell}^m(\unit{x})\right) } }\nonumber \\
&= \frac{ \sum\limits_{\ell,m}^{\LSlp-1} \sum\limits_{p,q}^{\LSlp-1} \overline{(g)_{\ell}^m} (g)_p^q K_{\ell m,p q} }{\sum\limits_{\ell,m}^{\LSlp-1} |(g)_{\ell}^m|^2},
\label{eq:lambda}
\end{align}
where
\begin{align}
K_{\ell m,p q} \dfn \int_R \overline{Y_{\ell}^m(\unit{x})} Y_p^q(\unit{x}) \, ds(\unit{x}),
\label{eq:K_elements}
\end{align}
and we have used the orthonormality of spherical harmonics on the sphere to get the final equality. Adopting the indexing introduced in \eqref{eq:flm_column}, we define an $\LSlp^2 \times \LSlp^2$ matrix $\B{K}$ with elements $K_{\ell m,pq}$ for $0 \le \ell,p < L_g, |m| \le \ell, |q| \le p$, and an $\LSlp^2 \times 1$ column vector $\B{g}$ with elements $(g)_{\ell}^m$ to rewrite \eqref{eq:lambda} in the matrix form as
\begin{align}
\lambda = \frac{ \B{g}^{\mathrm{H}} \B{K} \B{g} }{\B{g}^{\mathrm{H}} \B{g}},
\label{eq:lambda_matrix}
\end{align}
where $(\cdot)^{\mathrm{H}}$ represents conjugate transpose. Column vectors $\B{g}$ which render $\lambda$ in \eqref{eq:lambda_matrix} stationary are the solution to the following eigenvalue problem
\begin{align}
\B{K} \B{g} = \lambda \B{g}.
\label{eq:ev_problem}
\end{align}
From \eqref{eq:K_elements}, it can be seen that the matrix $\B{K}$ is Hermitian and positive definite, therefore, the eigenvalues $\lambda$ are real and eigenvectors $\B{g}$ are orthogonal\footnote{We choose the eigenvectors, $\B{g}$, to be orthonormal in this work.}. We index the eigenvalues~(and the associated eigenvectors) such that $1 > \lambda_1 \ge \lambda_2 \ge \ldots \ge \lambda_{\LSlp^2} > 0$. For each spectral domain eigenvector $\B{g}_{\alpha}$, associated with the eigenvalue $\lambda_{\alpha}$, we obtain a spatial eigenfunction given by
\begin{align}
g_{\alpha}(\unit{x}) = \sum_{\ell,m}^{\LSlp-1} (g_{\alpha})_{\ell}^m Y_{\ell}^m(\unit{x}), \quad 1 \le \alpha \le \LSlp^2,
\label{eq:spat_eig_func}
\end{align}
which are orthogonal over the spatial region $R$ and orthonormal over the sphere $\untsph$, i.e.,
\begin{align}
\begin{split}
& \innerp{g_{\alpha}}{g_{\beta}}_R = \B{g}_\alpha^{\mathrm{H}} \B{K} \,\B{g_\beta} =  \lambda_{\alpha} \delta_{\alpha,\beta}, \\
& \innerpS{g_{\alpha}}{g_{\beta}} =\B{g}_\alpha^{\mathrm{H}} \, \B{g}_\beta  =  \delta_{\alpha,\beta}.
\end{split}
\label{eqn:eigen_orthogo}
\end{align}
Set of spatial eigenfunctions, $g_{\alpha}(\theta,\phi),\,\alpha=1,2,\ldots,\LSlp^2$, serves as an alternative basis for the space of bandlimited signals, i.e, $\lsphL{\LSlp}$, and are referred to as Slepian functions. Consequently, any signal $h \in \lsphL{L_g}$ can be represented as
\begin{align}
h(\unit{x}) = \sum\limits_{\alpha = 1}^{\LSlp^2} (h)_{\alpha} g_{\alpha}(\unit{x}),\quad (h)_{\alpha} = \innerpS{h}{g_{\alpha}}= \B{g}_\alpha^{\mathrm{H}}\B{h}
\label{eq:Slep_expansion}
\end{align}
where $(h)_{\alpha},\, \alpha = 1,2,\ldots,\LSlp^2$, are called Slepian coefficients which constitute the Slepian domain representation of the signal $h$. \figref{fig:Slepian_functions} shows the first $12$ Slepian functions, bandlimited to $\LSlp = 32$ and computed over a spherical ellipse\footnote{We refer the reader to~\cite{Khalid:2013DSLSHT} for the definition of a spherical ellipse.}, rotated on the sphere by the Euler angles $\rho = (60^{\circ},90^{\circ},45^{\circ})$~(the ellipse is initially aligned with $x$-axis, having focus colatitude $\theta_c = 15^{\circ}$ and semi-arc-length of the semi-major axis $a = 20^{\circ}$).

As investigated in detail in~\cite{Simons:2006}, if most of the eigenvalues in \eqref{eq:ev_problem} are either nearly $1$ or nearly $0$~(suggesting maximal and minimal concentration for the corresponding eigenfunctions in the region $R$ respectively) with a sharp transition, then sum of the eigenvalues, called the spherical Shannon number, is a good measure of the number of well-optimally concentrated Slepian functions with in the region $R$. Denoted by $N_R$, the spherical Shannon number is given by~\cite{Simons:2006}
\begin{align}
N_R \dfn \sum_{\alpha=1}^{\LSlp^2} \lambda_{\alpha} = \mathrm{trace}(\B{K}) = \frac{A_{R}}{4\pi} \LSlp^2,
\label{eq:shannon_no}
\end{align}
where $A_R \dfn \|{1}\|_R$ is the surface area of the spatial region $R$. Hence, the first $N_R$ number of well-optimally concentrated Slepian functions in \eqref{eq:spat_eig_func}~(rounded to the nearest integer) form a (reduced) localized basis set for the accurate reconstruction and representation of bandlimited signals in the spatial region $R$.

\section{Spatial-Slepian Transform~(SST)}
\label{sec:SST}

In this section, we propose the spatial-Slepian transform~(SST) using the well-optimally concentrated Slepian functions. We show that the transform is invertible under some constraints, and establish that well-optimally localized rotated Slepian functions form a tight frame on the sphere. We also present a fast method for computing the proposed SST and carry out computational complexity analysis. We conclude this section with an illustration of SST on the Earth topography map using zonal Slepian functions computed over an axisymmetric polar cap region on the sphere.

\subsection{SST Formulation}
Slepian functions designed for bandlimit $\LSlp$ and spatial region $R$ on the sphere, i.e., $g_{\alpha}, \alpha = 1,2,\ldots,N_R$, can be used to define a new representation of signals on the sphere, which we refer to as the spatial-Slepian transform~(SST)\footnote{We use the term spatial-Slepian transform to differentiate it from Slepian transform which refers to the inner product between a signal and a Slepian function.} and define as
\begin{align}
F_{g_{\alpha}}(\rho) \dfn \innerpS{f}{(\Dp g_{\alpha})} = \intsph  f(\unit{x}) \overline{(\Dp g_{\alpha})(\unit{x})} \, ds(\unit{x}),
\label{eq:SST}
\end{align}
for a signal $f \in \lsphL{\Lf}$, where $\rho = (\varphi, \vartheta, \omega)$ is the $3$-tuple of the Euler angles, $\Dp \equiv \rotop$ is the rotation operator and $F_{g_{\alpha}} \in \SO$ is called the $\alpha^{\mathrm {th}}$ spatial-Slepian coefficient of the signal $f$. From its definition, we observe that spatial-Slepian transform probes the signal content by projecting it onto all possible rotated orientations of the well-optimally localized Slepian functions on the sphere, essentially spreading the signal in the so called joint spatial-Slepian domain. The extent of the spread of the signal in the joint spatial-Slepian domain, which is quantified by the number of spatial-Slepian coefficients, is specified by the rounded spherical Shannon number, and therefore, depends on the fractional surface area of the underlying region $R$ on the sphere and the bandlimit $\LSlp$ of the Slepian functions. In this context, we refer to $\alpha$ as the Slepian scale and $F_{g_{\alpha}}$ as the spatial-Slepian coefficient of Slepian scale $\alpha$.

Using the expansion of signals in \eqref{eq:f_expansion} and the spectral representation of the rotated signal in \eqref{eq:rotated_flms}, we can write the spatial-Slepian coefficient in \eqref{eq:SST} as
\begin{align}
F_{g_{\alpha}}(\rho) = \sum\limits_{\ell,m,m'}^{\min\{\Lf-1,\LSlp-1\}} (f)_{\ell}^{m} \overline{(g_{\alpha})_{\ell}^{m'}} \, \overline{D^{\ell}_{m,m'}(\rho)},
\label{eq:SST2}
\end{align}
where we have used orthonormality of spherical harmonics on the sphere to obtain the final expression.

\subsection{Inverse SST}
Since, the spatial-Slepian coefficient $F_{g_{\alpha}}(\rho)$ can be expressed as a weighted sum of conjugate of Wigner-$D$ functions, we define the Fourier representation of $F_{g_{\alpha}}(\rho)$ as
\begin{align}
(F_{g_{\alpha}})^{\ell}_{m,m'} \! \dfn \left(\frac{2\ell+1}{8\pi^2}\right) \!\! \innerpSO{F_{g_{\alpha}}}{\overline{D_{m,m'}^{\ell}}} \!\!\!\!\!\! = (f)_{\ell}^m \overline{(g_\alpha)_{\ell}^{m'}}
\label{eq:SST_spectral}
\end{align}
for $0 \le \ell, |m|, |m'| \le \min\{\Lf-1,\LSlp-1\}$, where we have used \eqref{eq:inner_prod_SO3} to obtain the final result. Hence, we can recover the spectral coefficients of the original signal $f$ as
\begin{align}
(f)_{\ell}^m &= \left(\frac{2\ell+1}{8\pi^2}\right) \frac{ \innerpSO{F_{g_{\alpha}}} {\overline{D_{m,m'}^{\ell}}} } { \overline{(g_\alpha)_{\ell}^{m'}} } \nonumber \\
&= \left(\frac{2\ell+1}{8\pi^2}\right) \frac{ \displaystyle \intSO F_{g_{\alpha}}(\rho) D_{m,m'}^{\ell}(\rho) \, d\rho } { \overline{(g_\alpha)_{\ell}^{m'}} }
\label{eq:inv_SST}
\end{align}
for $0 \le \ell, |m|, |m'| \le \min\{\Lf-1,\LSlp-1\}$. From \eqref{eq:inv_SST}, we note that the proposed spatial-Slepian transform is invertible only if the spherical harmonic coefficients of the Slepian functions, $(g_{\alpha})_{\ell}^{m'}$, are non-zero for all degrees $0 \le \ell \le \min\{\Lf-1,\LSlp-1\}$ and at least one order $-\ell \le m' \le \ell$.

\begin{remark}
	For the case where $\Lf > \LSlp$, the inverse spatial-Slepian transform cannot recover all of the spectral coefficients of the signal $f$. On the other hand if $\Lf < \LSlp$,  the Slepian functions are under-utilized in spatially localizing the signal $f$. Therefore, in this work, we assume that $\Lf = \LSlp$, so that not only the Slepian functions are fully utilized, signal $f$ is also perfectly recovered from its spatial-Slepian representation.
	\label{rem:Rem1}
\end{remark}

\subsection{Tight frame}
A sequence of functions $\{\varphi_n\}_{n \in \N}$ in a Hilbert space $\mathcal{H}$ is called a frame if there exists $0 \le A \le B < \infty$ such that
\begin{align}
A \norm{f}^2 \le \sum_{n \in \N} \left|\innerp{f}{\varphi_n}_{\mathcal{H}}\right|^2 \le B \norm{f}^2, \qquad \forall \, f \in \mathcal{H},
\label{eq:frame_cond}
\end{align}
where $\innerp{\cdot}{\cdot}_{\mathcal{H}}$ is the inner product defined for the Hilbert space $\mathcal{H}$ and $A$, $B$ are called lower and upper frame bounds respectively. If $A = B$, then the sequence of functions $\{\varphi_n\}_{n \in \N}$ in \eqref{eq:frame_cond} is called a tight frame. For a tight frame, we have
\begin{align}
\norm{f}^2 = \frac{1}{A} \sum_{n \in \N} \left|\innerp{f}{\varphi_n}_{\mathcal{H}}\right|^2.
\label{eq:tight_frame_cond}
\end{align}

Consider the Slepian functions $g_{\alpha}, \alpha = 1,2,\ldots, N_R$, which are used to obtain the spatial-Slepian coefficients $F_{g_{\alpha}}$ in \eqref{eq:SST}. Then, we can write
\begin{align}
&\sum_{\alpha=1}^{N_R} \intSO \left|F_{g_{\alpha}}(\rho)\right|^2 d\rho = \sum_{\alpha=1}^{N_R} \intSO \left|\innerpS{f}{(\Dp g_{\alpha})}\right|^2 d\rho \nonumber \\
&= \sum_{\alpha=1}^{N_R} \sum_{\ell,m}^{\Lf-1} |(f)_{\ell}^m|^2 \sum_{s,t,t'}^{\LSlp-1} \overline{(g_{\alpha})_s^{t'}} \sum_{w'=-s}^s \, (g_{\alpha})_s^{w'} \times \nonumber \\
& \qquad \qquad \qquad \qquad \qquad \intSO \overline{D^s_{t,t'}(\rho)}  D^s_{t,w'}(\rho) d\rho  \nonumber \\
&= \sum_{\alpha=1}^{N_R} \left[ \sum_{s,t}^{\LSlp-1} \left(\frac{8\pi^2}{2s+1}\right)\sum_{t'=-s}^s \, |(g_{\alpha})_s^{t'}|^2 \right] \sum_{\ell,m}^{\Lf-1} |(f)_{\ell}^m|^2,
\end{align}
where we have used orthonormality of spherical harmonics on the sphere and orthogonality of Wigner-$D$ functions on the $\SO$ rotation group to obtain the final result. Hence,
\begin{align}
\norm{f}^2 = \sum_{\ell,m}^{\Lf-1} |(f)_{\ell}^m|^2 = \frac{\sum\limits_{\alpha=1}^{N_R} \displaystyle\intSO \left|\innerpS{f}{(\Dp g_{\alpha})}\right|^2}{\sum\limits_{\alpha=1}^{N_R} \left[ \sum\limits_{s,t,t'}^{\LSlp-1} \left(\frac{8\pi^2}{2s+1}\right) \, |(g_{\alpha})_s^{t'}|^2 \right]},
\label{eq:tight_frame_proof}
\end{align}
which shows that the well-optimally localized rotated Slepian functions, $(\Dp g_{\alpha})$, $\alpha = 1,2,\ldots,N_R$, form a tight frame for the Hilbert space of bandlimited functions $\lsphL{\LSlp}$.

\subsection{Fast Computation of Spatial-Slepian Transform}
Using the definition of Wigner-$D$ functions in \eqref{eq:WignerD}, we can write the spatial-Slepian coefficients in \eqref{eq:SST2} as
\begin{align}
&F_{g_{\alpha}}(\varphi, \vartheta, \omega) = \sum_{\ell,m,m'}^{\Lf - 1} (f)_{\ell}^m \, \overline{(g_{\alpha})_{\ell}^{m'}} \, e^{im\varphi} \, \overline{d^{\ell}_{m,m'}(\vartheta)} \, e^{im'\omega} \nonumber \\
&= \sum_{\ell=0}^{\Lf - 1} \sum_{m=-\ell}^{\ell} (f)_{\ell}^m \, \overline{(g_{\alpha})_{\ell}^{m'}} \, e^{im\varphi} \times \nonumber \\
& \qquad \sum_{m'=-\ell}^{\ell} i^{m'-m} e^{im'\omega} \sum_{m''=-\ell}^{\ell} \Delta^{\ell}_{m'',m} \,\Delta^{\ell}_{m'',m'} e^{im''\vartheta},
\label{eq:SST3}
\end{align}
where $\Delta^{\ell}_{m,m'} \dfn d^{\ell}_{m,m'}(\pi/2)$ and we have used the following expansion for Wigner-$d$ functions~\cite{Kennedy-book:2013},
\begin{align}
d^{\ell}_{m,m'}(\vartheta) = i^{m-m'} \sum_{m''=-\ell}^{\ell} \Delta^{\ell}_{m'',m} \Delta^{\ell}_{m'',m'} e^{-im''\vartheta}.
\label{eq:Wigner_d_expansion}
\end{align}
By rearranging the summations in \eqref{eq:SST3}, we can rewrite the spatial-Slepian coefficient as
\begin{align}
F_{g_{\alpha}}(\rho) &= \sum_{m,m',m''=-(\Lf-1)}^{\Lf-1} C_{m,m',m''} e^{i(m\varphi+m''\vartheta+m'\omega)},
\label{eq:SST_fast}
\end{align}
where
\begin{align}
C_{m,m',m''} = i^{m'-m} \!\!\!\!\!\!\!\!\!\! \sum_{\ell = \max\{|m|,|m'|,|m''|\}}^{\Lf-1} \!\!\!\!\!\!\!\! (f)_{\ell}^m \, \overline{(g_{\alpha})_{\ell}^m} \, \Delta^{\ell}_{m'',m} \Delta^{\ell}_{m'',m'}.
\label{eq:C_mm'm''}
\end{align}
The expression in \eqref{eq:SST_fast} is a simple rearrangement of the initial expression in \eqref{eq:SST2} and hence, is not more efficient. However, the presence of complex exponential functions in \eqref{eq:SST_fast} facilitates the use of the fast Fourier transform~(FFT) algorithm to compute the spatial-Slepian coefficient efficiently. Wigner-$d$ functions $\Delta^{\ell}_{m,m'}$ can be computed using either the recursive relations given in~\cite{Trapani:2006} or the recursion proposed in~\cite{Risbo:1996}, both of which are stable up to very large degrees.

\subsubsection{Computational Complexity Analysis}
We observe that spatial-Slepian transform in \eqref{eq:SST_fast} requires the computation of the coefficients $C_{m,m',m''}$ over the three dimensional space of orders $m$, $m'$ and $m''$. Coefficients $C_{m,m',m''}$ in turn require a single summation over the degree $\ell$ for each $m$, $m'$, $m''$. As a result, the overall complexity of computing $C_{m,m',m''}$ scales as $O(\Lf^4)$ with bandlimit $\Lf$. We note that Wigner-$d$ functions $\Delta^{\ell}_{m,m'}$ do not depend on either the signal or Slepian functions and hence, can be computed in $O(\Lf^3)$ using the recursion in~\cite{Trapani:2006}. However, we compute $\Delta^{\ell}_{m,m'}$ on-the-fly to minimize storage requirements and note that this does not change the overall complexity of $O(\Lf^4)$ for computing the coefficients $C_{m,m',m''}$. Computational complexity of the three dimensional fast Fourier transform scales as $O(\Lf^3\log_2\Lf)$ with bandlimit $\Lf$. Hence, the overall complexity for computing the spatial-Slepian coefficient in \eqref{eq:SST_fast} is governed by the coefficients $C_{m,m',m''}$, and is given by $O(\Lf^4)$ for a fixed Slepian scale $\alpha$, and $O(N_R\Lf^4)$ for all Slepian scales, i.e., $\alpha = 1,2,\ldots,N_R$.

\begin{figure}[!t]
	\centering
	\includegraphics[width=0.48\textwidth]{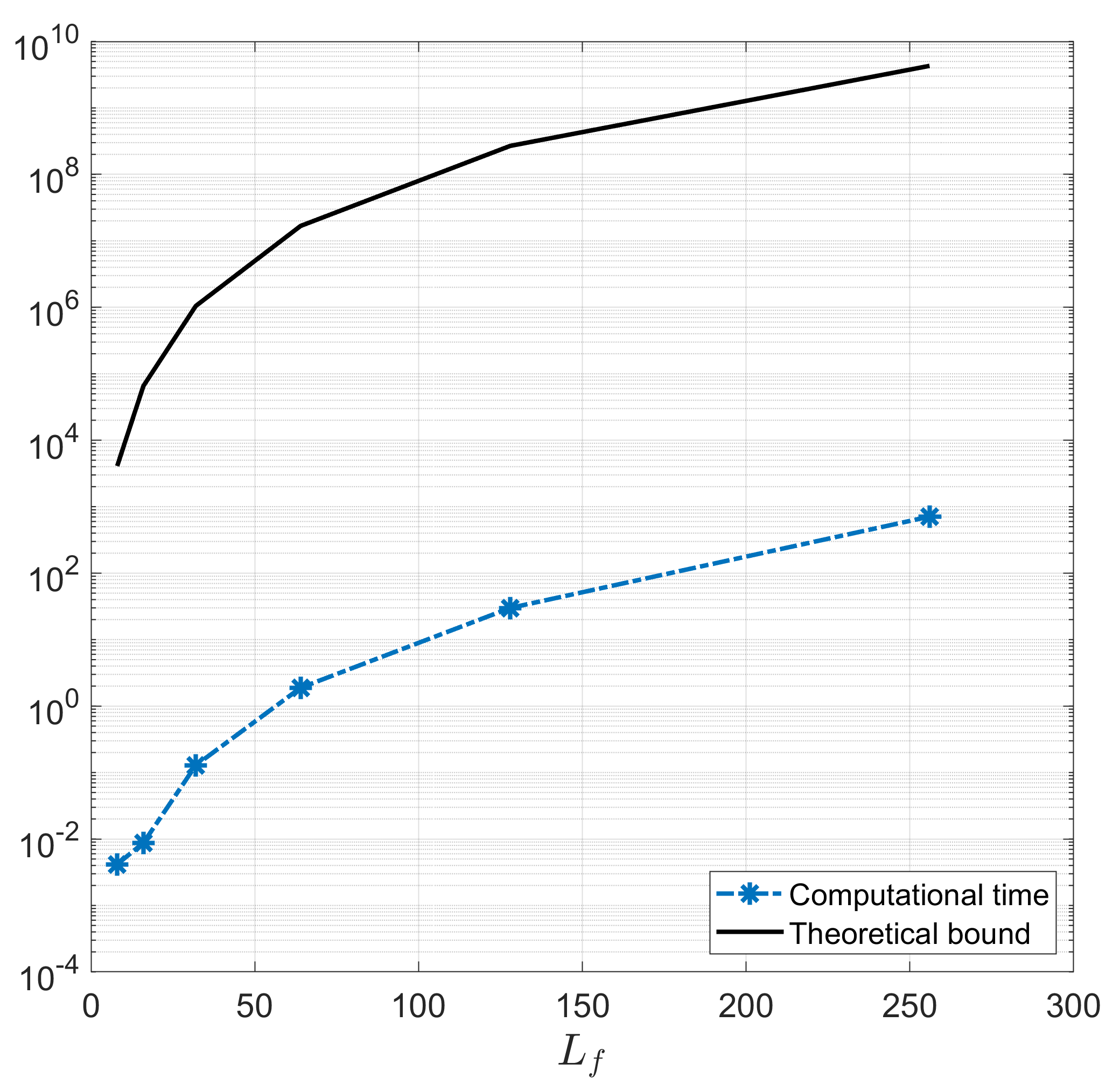}
	\caption{Computational complexity analysis of the spatial-Slepian transform for a test signal using one of the Slepian functions, at Slepian scale $\alpha = 1$, computed over a spherical ellipse, which is aligned with $x$-axis, having focus colatitude $\theta_c = 15^{\circ}$ and semi-arc-length of the semi-major axis $a = 20^{\circ}$.}
	\label{fig:comp_time}	
\end{figure}

We validate the computational complexity of the spatial-Slepian transform using one of the Slepian functions~(at Slepian scale $\alpha = 1$), computed over a spherical ellipse which is aligned with $x$-axis, having focus colatitude $\theta_c = 15^{\circ}$ and semi-arc-length of the semi-major axis $a = 20^{\circ}$. Spatial-Slepian transform is computed for a test signal which is generated in the spectral domain such that the spectral coefficients are complex, with real and imaginary parts uniformly distributed in the interval $(0,1)$. The experiment is performed in $\matlab$, running on a $2.2$ GHz Intel Core i$7$ processor with $16$ GB RAM. We record the time required to compute the spatial-Slepian coefficients at different values of the bandlimit $\Lf$ and plot it in \figref{fig:comp_time}, where we have also shown the theoretical bound which scales as $O(\Lf^4)$. As expected, the results in \figref{fig:comp_time} corroborate the theoretically established bound on the computational complexity of the spatial-Slepian transform.

\subsection{SST using Zonal Slepian functions over Axisymmetric North Polar Cap Region}
\begin{figure*}[!t]
	\centering
	\subfloat[$f(\unit{x})$]{
		\includegraphics[width=0.15\textwidth]{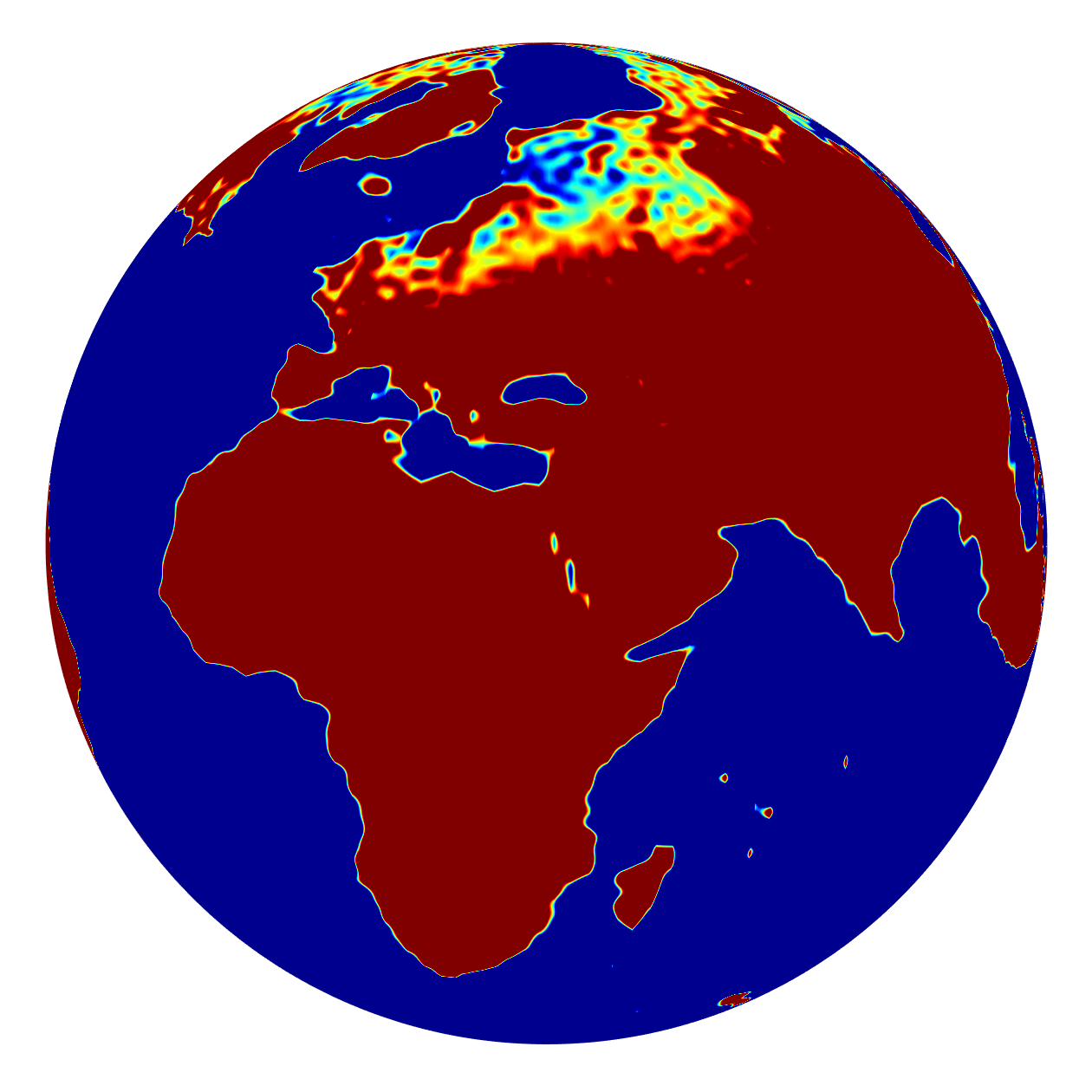}}\hfil
	\subfloat[$F_{g_1}(\unit{x})$]{
		\includegraphics[width=0.15\textwidth]{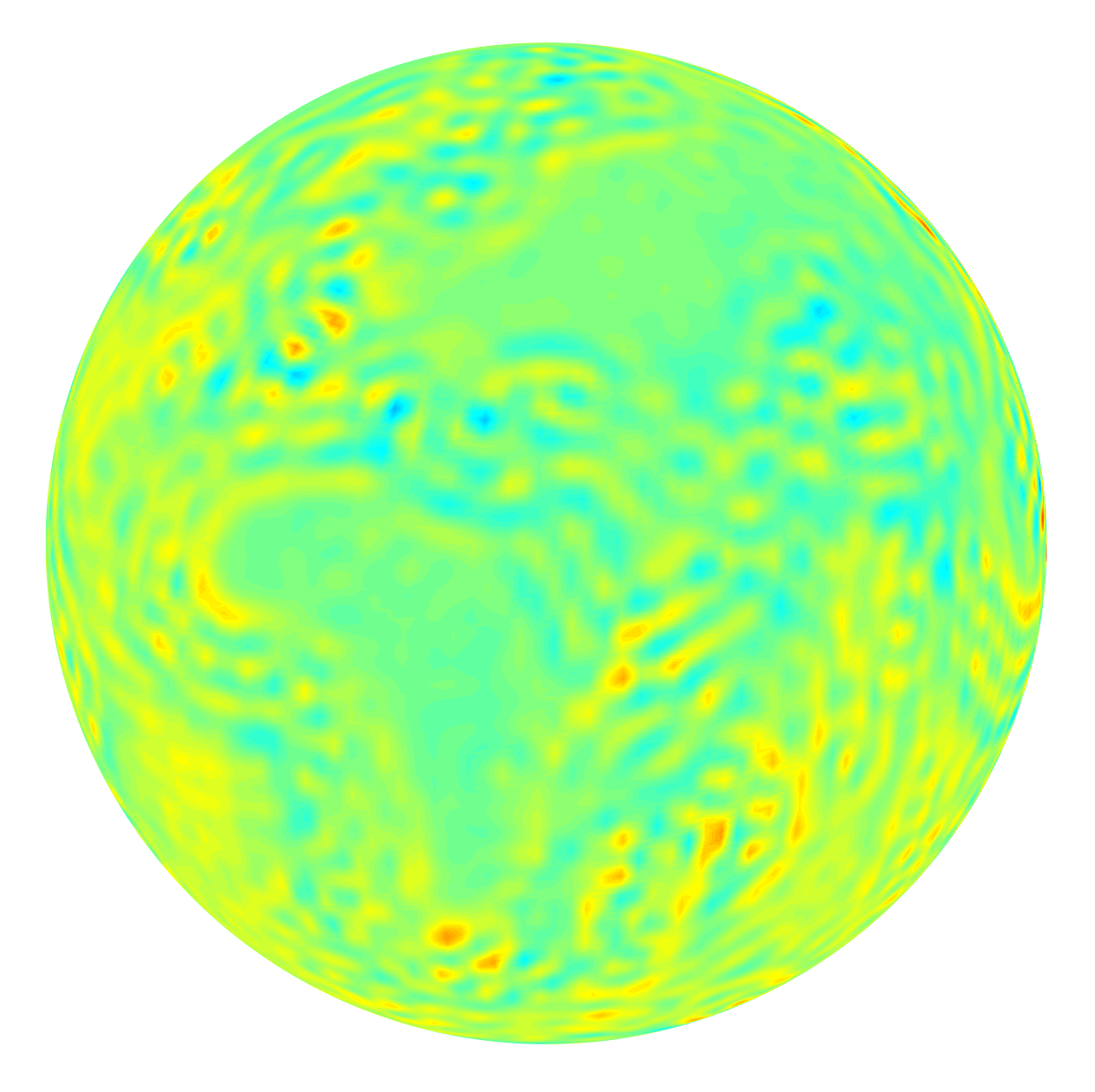}}\hfil
	\subfloat[$F_{g_2}(\unit{x})$]{
		\includegraphics[width=0.15\textwidth]{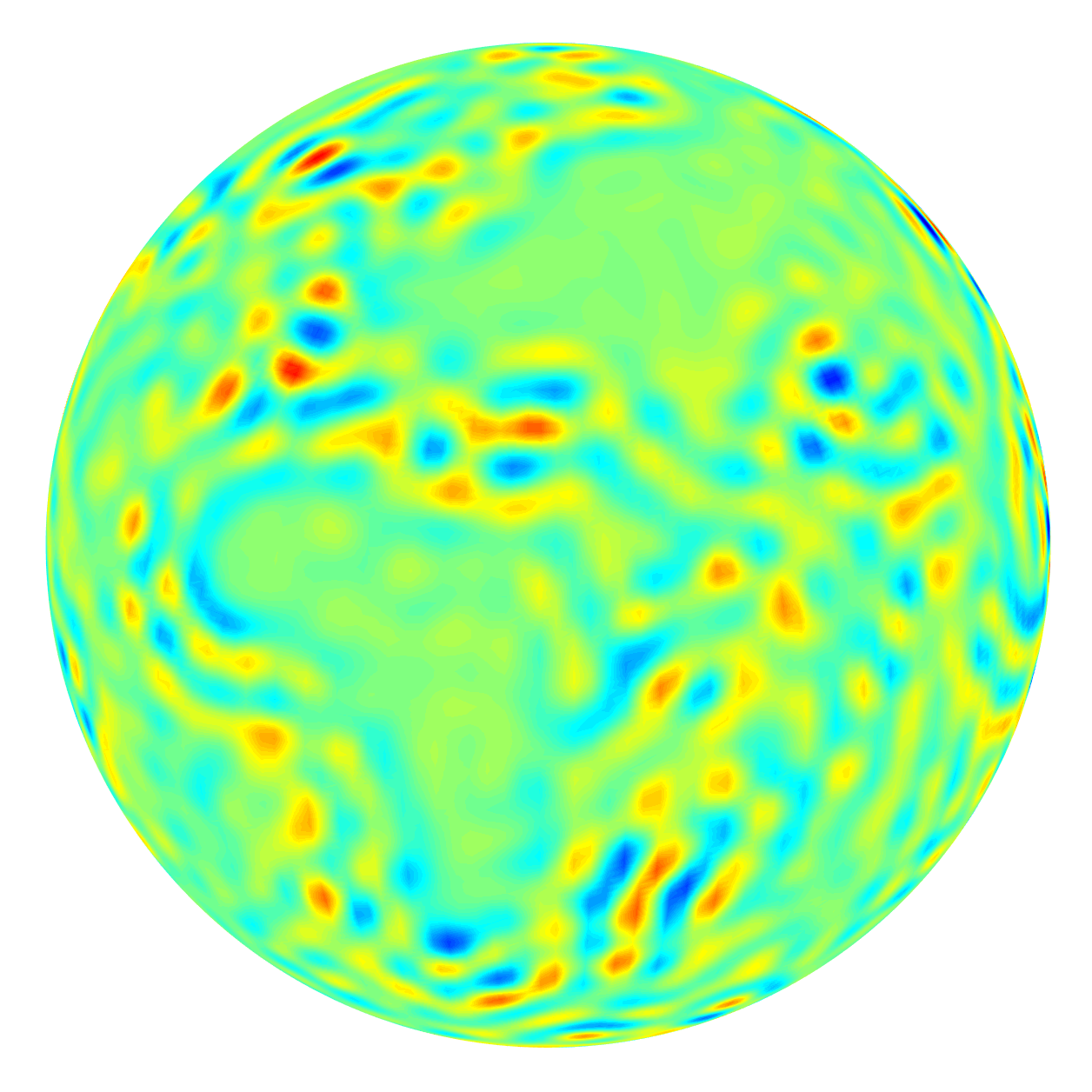}}\hfil
	\subfloat[$F_{g_3}(\unit{x})$]{
		\includegraphics[width=0.15\textwidth]{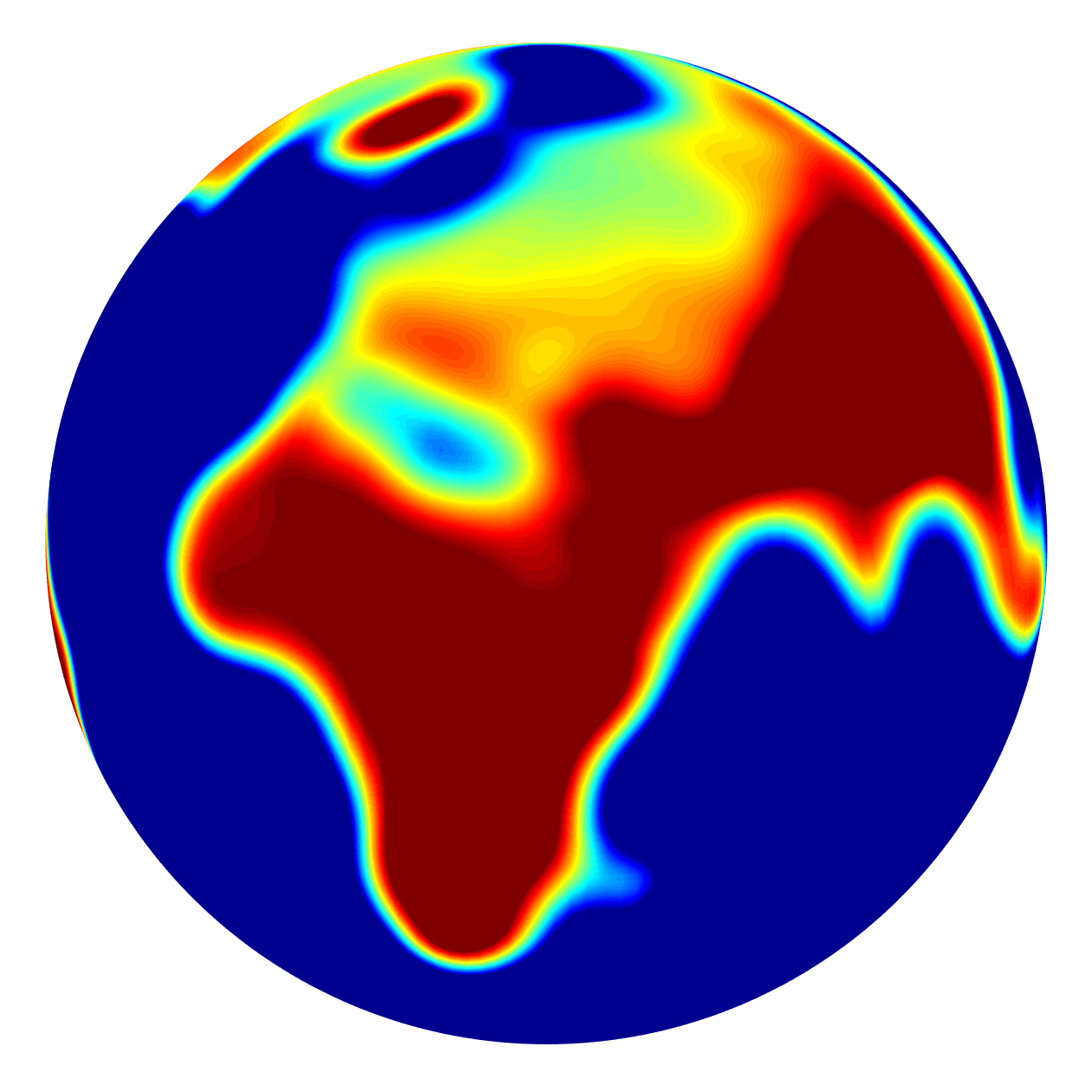}}\hfil	
	\subfloat[$F_{g_4}(\unit{x})$]{
		\includegraphics[width=0.15\textwidth]{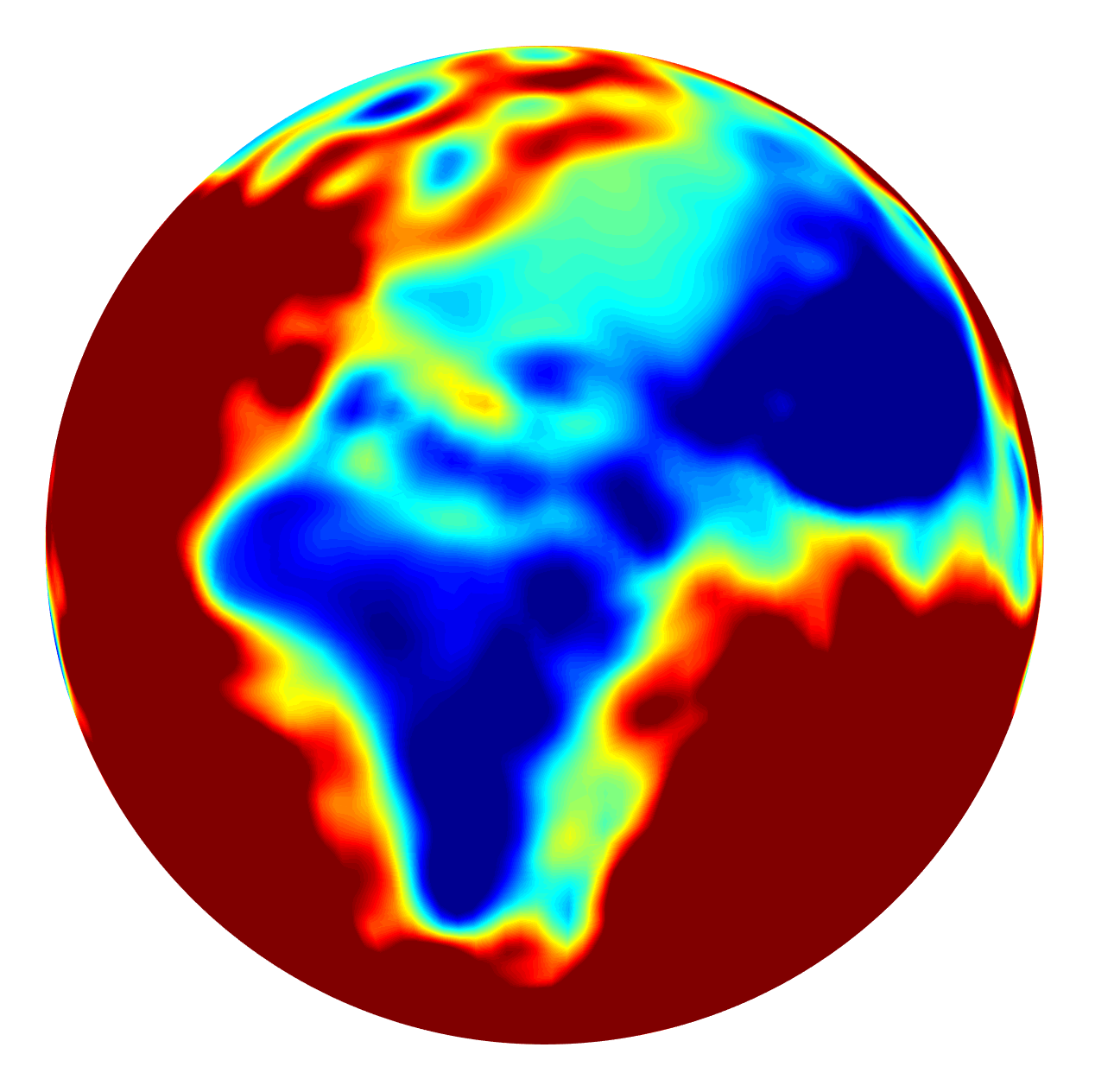}}\hfil	
	\subfloat[$F_{g_5}(\unit{x})$]{
		\includegraphics[width=0.15\textwidth]{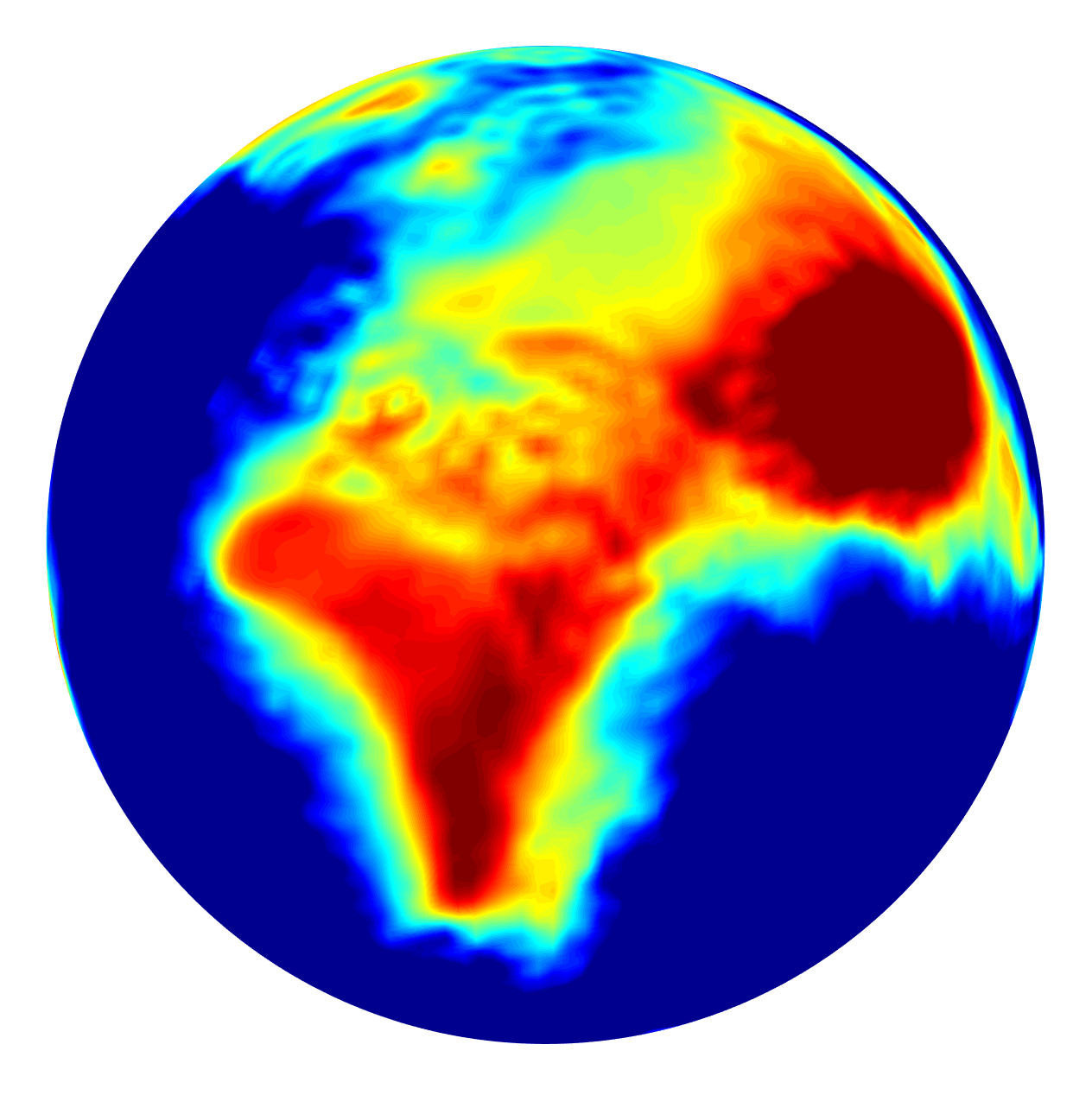}}\hfil
	
	\subfloat[$F_{g_6}(\unit{x})$]{
		\includegraphics[width=0.15\textwidth]{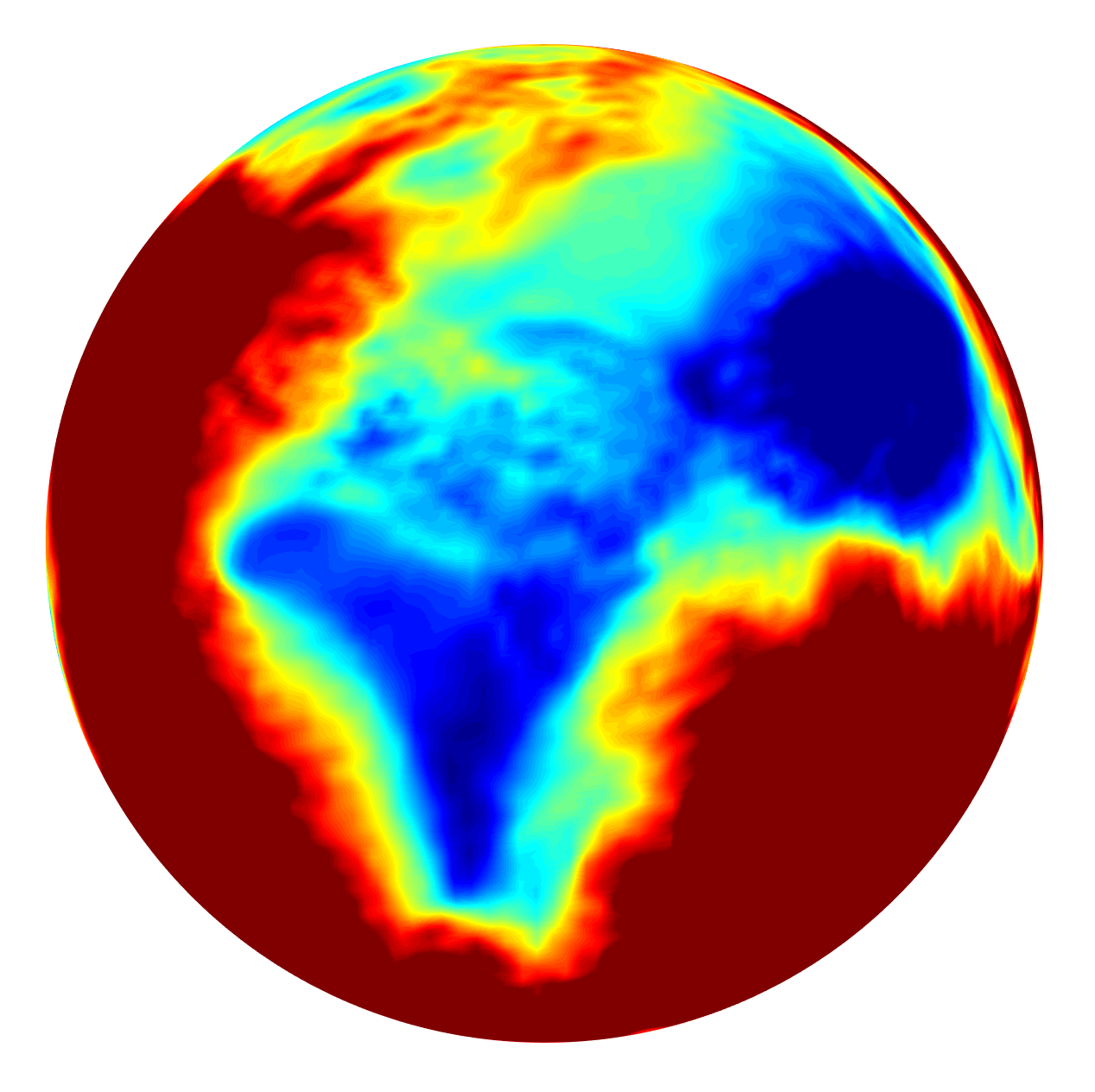}}\hfil	
	\subfloat[$F_{g_7}(\unit{x})$]{
		\includegraphics[width=0.15\textwidth]{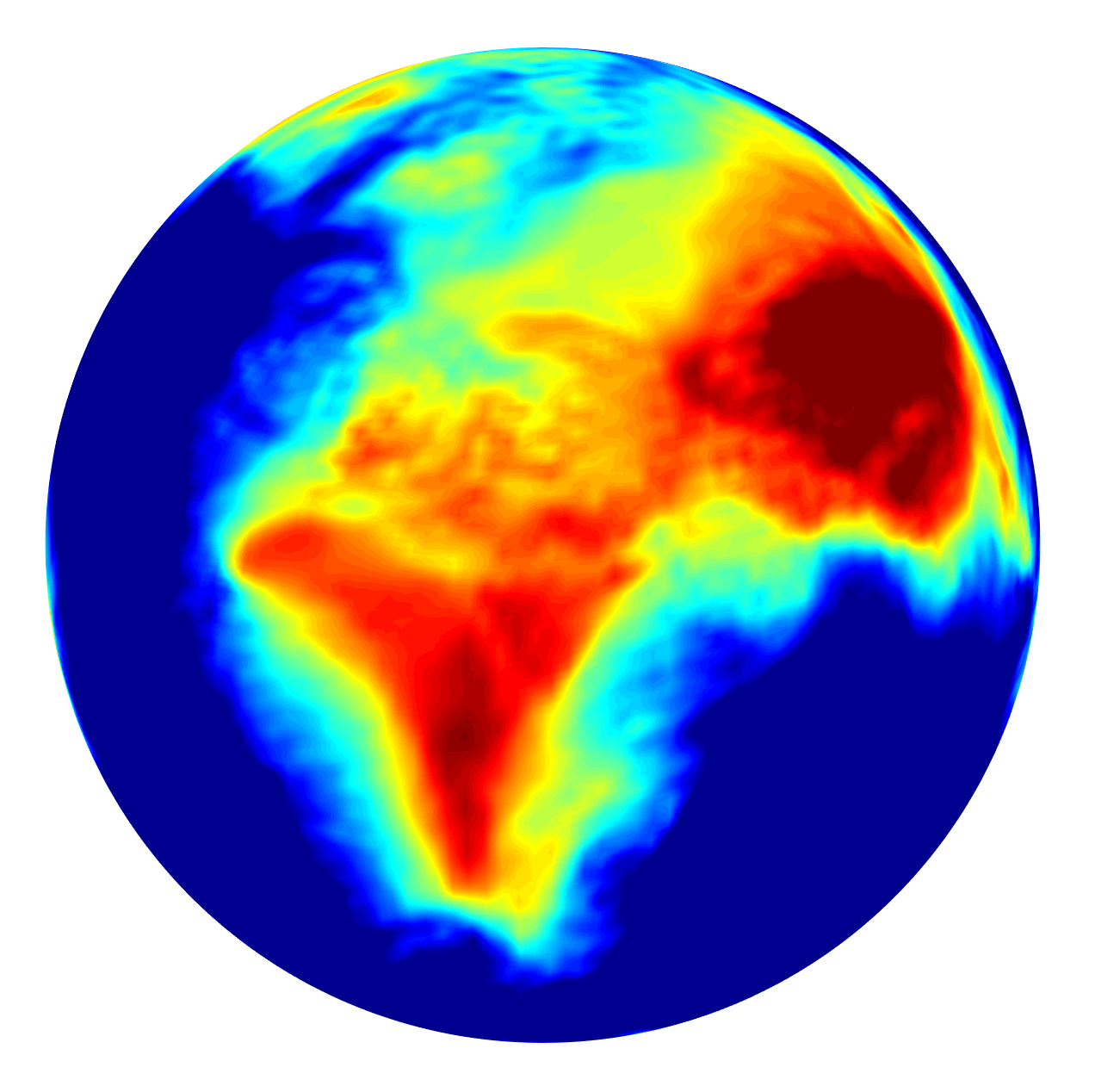}}\hfil
	\subfloat[$F_{g_8}(\unit{x})$]{
		\includegraphics[width=0.15\textwidth]{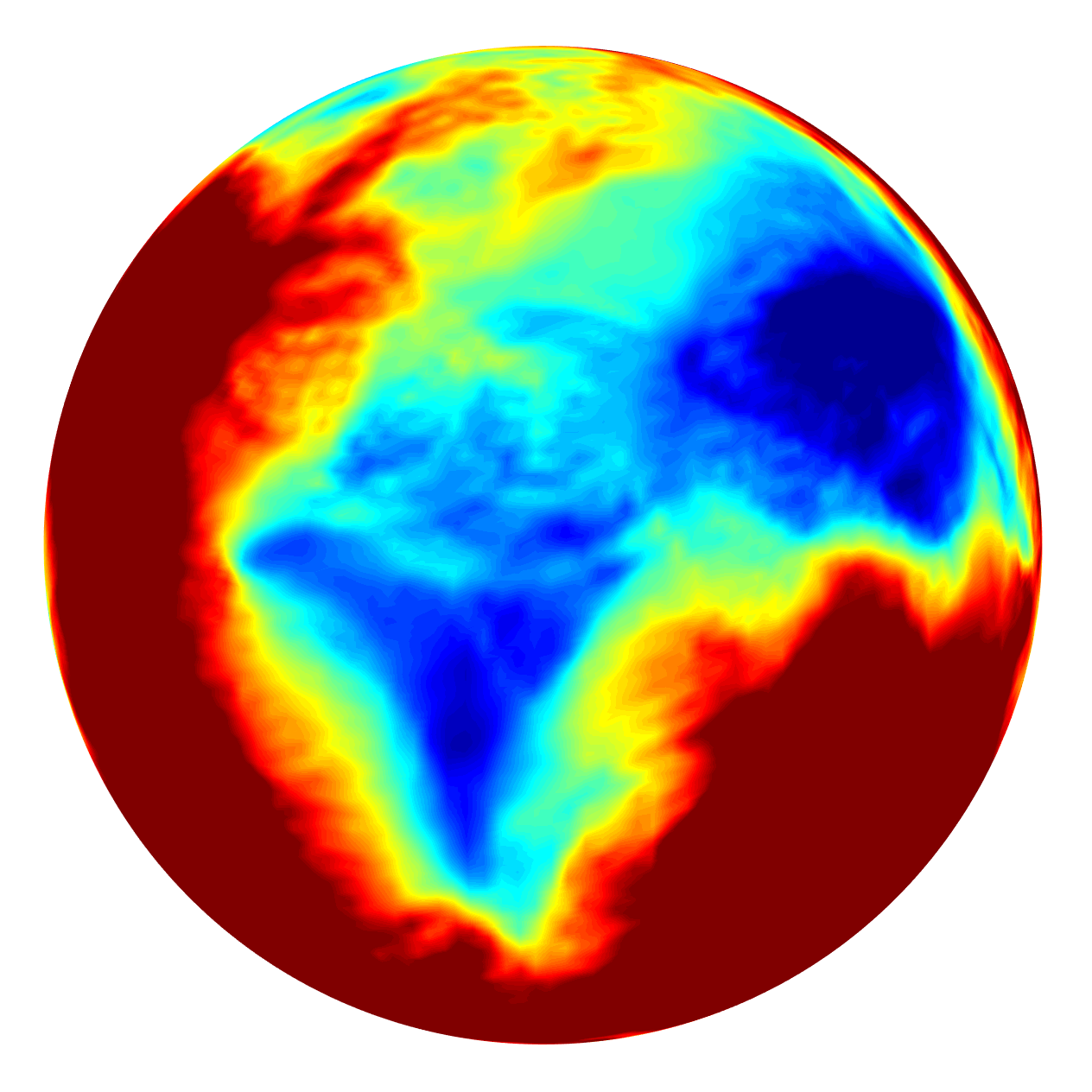}}\hfil
	\subfloat[$F_{g_9}(\unit{x})$]{
		\includegraphics[width=0.15\textwidth]{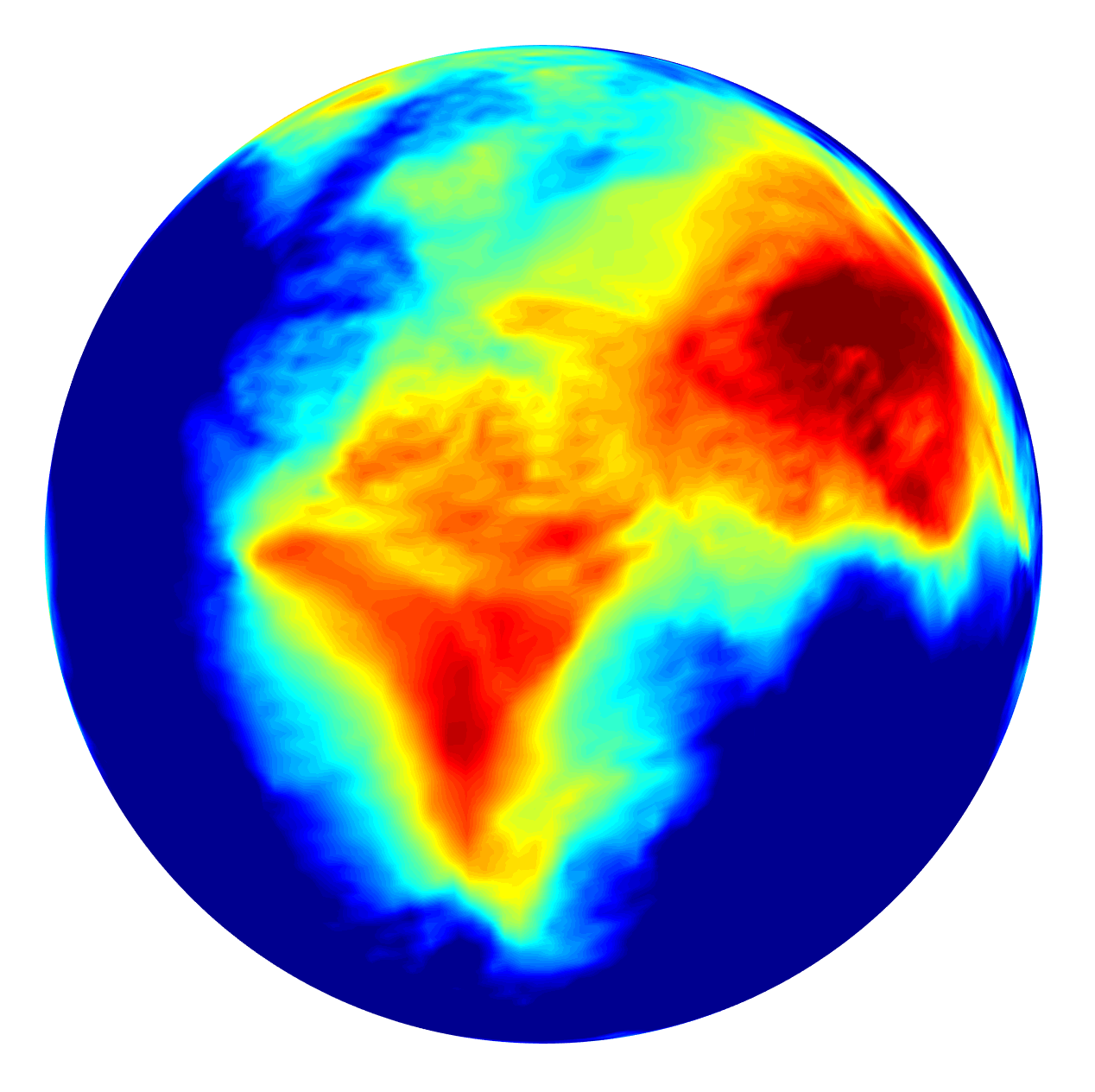}}\hfil
	\subfloat[$F_{g_{10}}(\unit{x})$]{
		\includegraphics[width=0.15\textwidth]{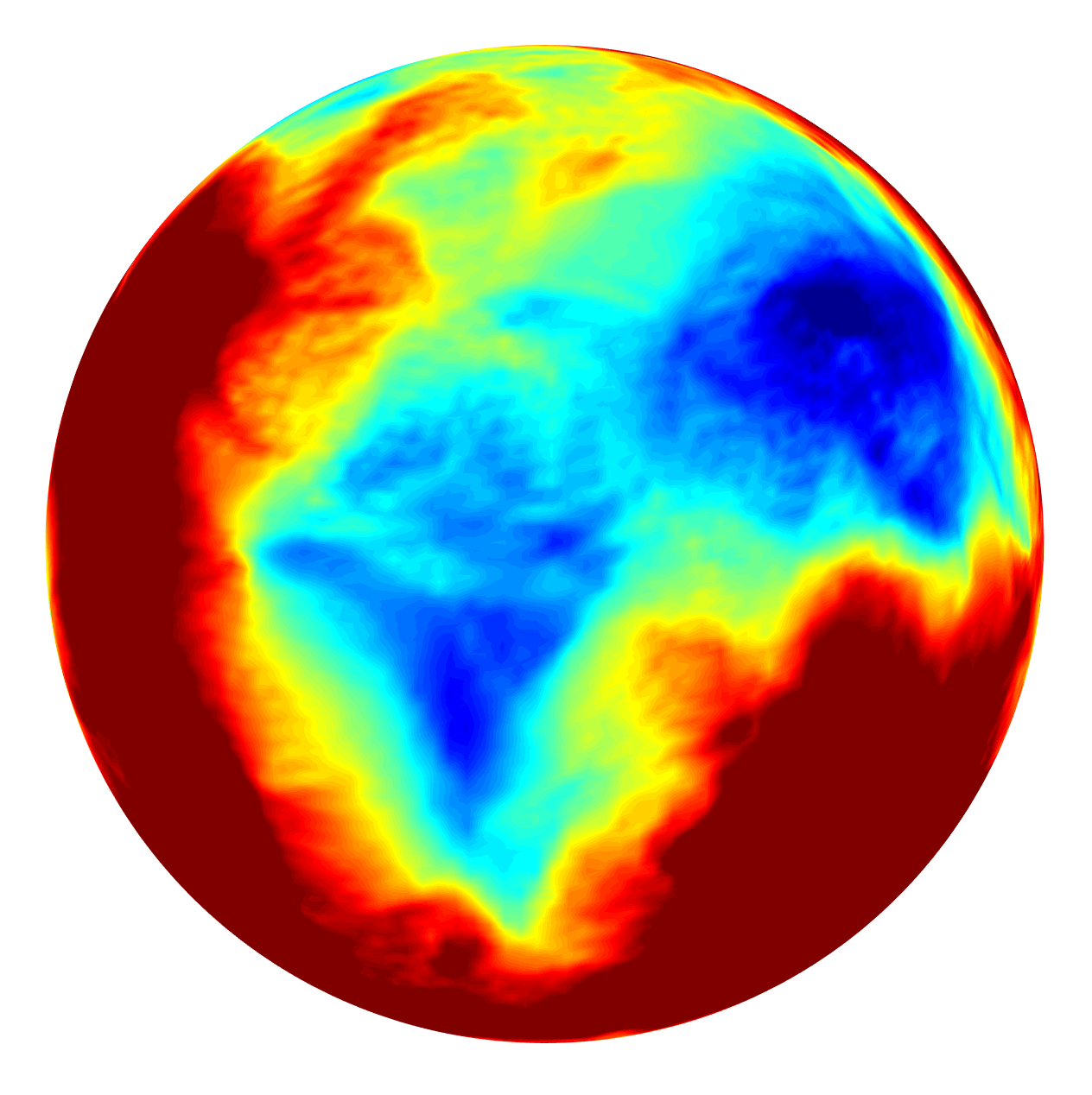}}\hfil
	\subfloat[$F_{g_{11}}(\unit{x})$]{
		\includegraphics[width=0.15\textwidth]{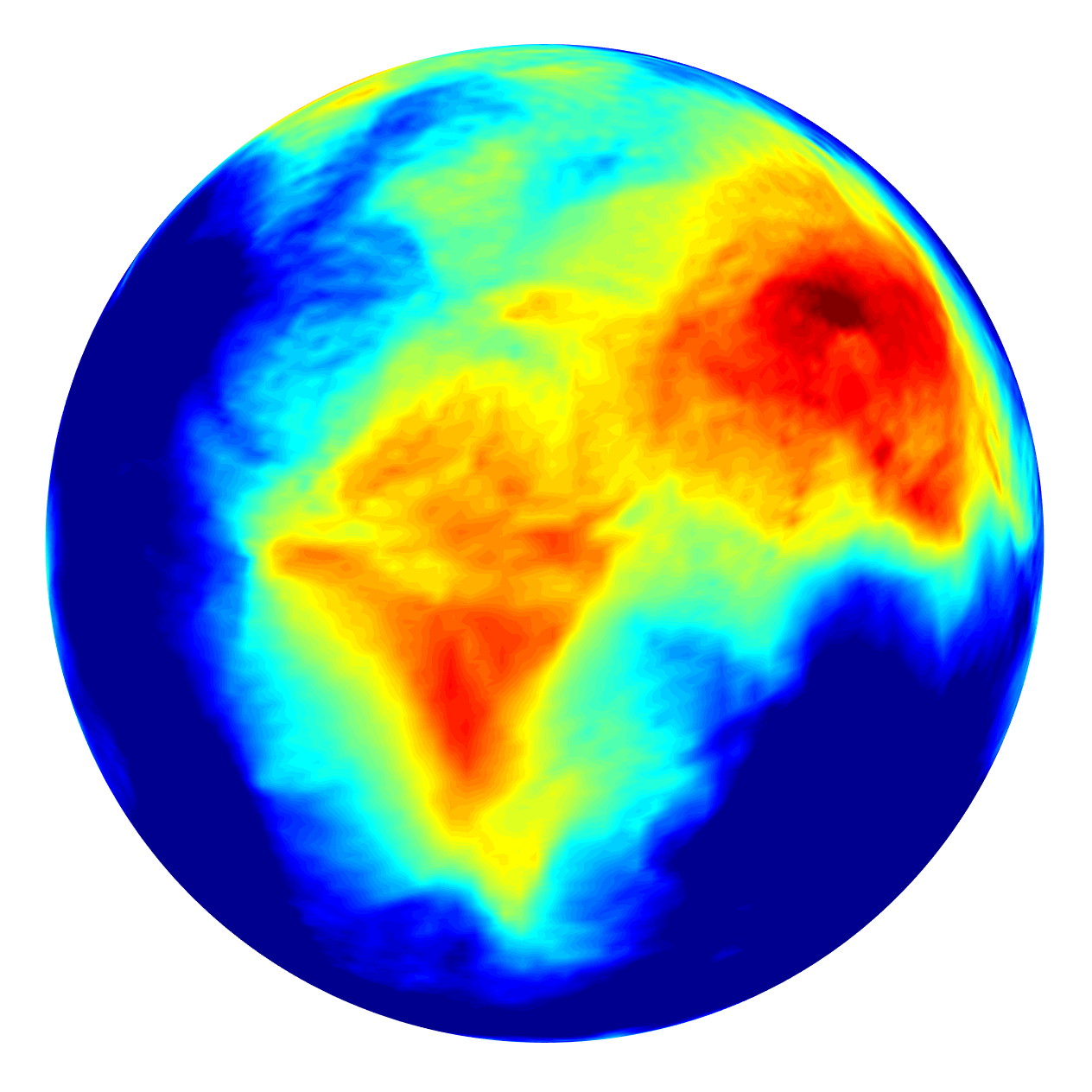}}\hfil
	
	\subfloat{
		\includegraphics[width=\textwidth]{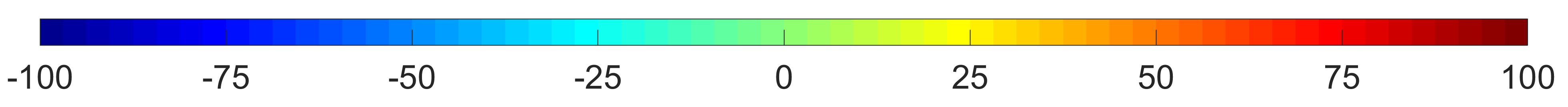}}\hfil
	\caption{Earth topography map and $N_{\pr,0} \sim 11$ spatial-Slepian coefficients for the Earth topography map at bandlimit $L_f = 128$, using zonal Slepian functions computed over axisymmetric north polar cap region of polar cap angle $\pr = 15^{\circ}$.}
	\label{fig:Earth_SST}
\end{figure*}
Axisymmetric north polar cap region is specified by a polar cap angle $\pr$, and is defined as $\{\unit{x}(\theta,\phi) \in \R^3 : |\unit{x}| = 1, 0 \le \theta \le \pr, 0 \le \phi < 2\pi\}$. Slepian spatial-spectral concentration problem for polar cap regions has been investigated and analytically solved in~\cite{Simons:2006}. The resulting Slepian functions are axisymmetric, i.e., $g(\theta,\phi) = g(\theta)$. In particular, we use the order $m = 0$ Slepian functions called zonal Slepian functions to compute spatial-Slepian transform. Spectral representation for zonal Slepian functions becomes
\begin{align}
(g_{\alpha})_{\ell}^m = (g_{\alpha})_{\ell}^0 \delta_{m,0},
\label{eq:flm_zonal_Slepian}
\end{align}
and the spherical Shannon number is given by~\cite{Wieczorek:2005}
\begin{align}
N_{\pr,0} = L \frac{\pr}{\pi}.
\label{eq:shannon_no_zonal}
\end{align}
Using \eqref{eq:flm_zonal_Slepian}, we can write the rotated signal $(\Dp g_{\alpha})(\theta,\phi)$ in \eqref{eq:SST} as
\begin{align}
(\Dp g_{\alpha})(\unit{x}) &= \sum_{\ell,m}^{L-1} \sqrt{\frac{4\pi}{2\ell+1}} \overline{Y_{\ell}^m(\vartheta,\varphi)} (g_{\alpha})_{\ell}^0 Y_{\ell}^m(\unit{x}),
\label{eq:rot_sig_axisym}
\end{align}
where we have used the fact that for $m' = 0$, the first rotation by $\omega$ around $z$-axis has no effect and can be taken to be $0$, along with the following relation~\cite{Kennedy-book:2013} to obtain the final result
\begin{align}
D^{\ell}_{m,0}(\varphi,\vartheta,0) = \sqrt{\frac{4\pi}{2\ell+1}} \overline{Y_{\ell}^m(\vartheta,\varphi)}.
\label{eq:WignerD_Ylm}
\end{align}
Therefore, using the orthonormality of spherical harmonics on the sphere, SST in \eqref{eq:SST} can be rewritten for zonal Slepian functions over an axisymmetric polar cap region as
\begin{align}
F_{g_{\alpha}}(\rho) &= \innerpS{f}{(\Dp g_{\alpha})} \nonumber \\
&= \sum\limits_{\ell,m}^{L-1} \sqrt{\frac{4\pi}{2\ell+1}} (f)_{\ell}^{m} \overline{(g_{\alpha})_{\ell}^0} \, Y_{\ell}^m(\vartheta,\varphi) = F_{g_{\alpha}}(\vartheta,\varphi).
\label{eq:SST_axisym}
\end{align}
Note that the spatial-Slepian coefficient, in this case, is a signal on the sphere $\untsph$, with spherical harmonic coefficients given by
\begin{align}
(F_{g_{\alpha}})_{\ell}^m = \innerpS{F_{g_{\alpha}}}{Y_{\ell}^m} = \sqrt{\frac{4\pi}{2\ell+1}} (f)_{\ell}^{m} \overline{(g_{\alpha})_{\ell}^0}.
\label{eq:SST_spectral_axisym}
\end{align}
As a result, signal $f$ can be reconstructed perfectly from the spatial-Slepian coefficients as
\begin{align}
f(\theta,\phi) = \sum_{\ell,m}^{L-1} \left[ \sqrt{\frac{2\ell+1}{4\pi}} \frac{ \innerpS{F_{g_{\alpha}}} {Y_{\ell}^m} }{\overline{(g_{\alpha})_{\ell}^0}} \right] Y_{\ell}^m(\theta,\phi),
\label{eq:f_from_SST_axisym}
\end{align}
for $\overline{(g_\alpha)_{\ell}^0} \ne 0, \forall \, \ell < \Lf$. We use the Earth topography map\footnote{http://geoweb.princeton.edu/people/simons/software.html}, bandlimited to degree $\Lf = 128$, for the computation of spatial-Slepian transform using zonal Slepian functions computed over the axisymmetric north polar cap region with polar cap angle $\pr = 15^{\circ}$. \figref{fig:Earth_SST} shows the spatial-Slepian coefficients for the first $N_{\pr,0} \sim 11$ Slepian scales, along with the Earth topography map.

\section{Localized Variation Analysis}
\label{sec:LVA}

As discussed in \secref{sec:slep}, Slepian functions form an alternative basis set for the representation of bandlimited signals on the sphere and the well-optimally concentrated Slepian basis functions form a (reduced)~localized basis set for the accurate representation and reconstruction of bandlimited signals over a region on the sphere. Hence, this reduced basis can prove to be an invaluable tool for probing the contents of any signal which is localized with in a region on the sphere. In this context, we  present an application of the spatial-Slepian transform, utilizing it for detecting hidden variations in a signal, which are localized with in an unknown region on the sphere. The objective is to detect the presence of these variations along with an estimate of the underlying region that these variations are localized with in. In the remainder of this section, we setup the problem of localized variation analysis and use a toy example for illustration. We compare the results obtained using the spatial-Slepian transform with those obtained from the wavelet transform~\cite{McEwen:2018}, and show that spatial-Slepian transform performs better by achieving a better estimate of the underlying region of localized variations.

\subsection{Problem Statement}
Let $b(\unit{x})$ be an unknown signal on the sphere, called the background source signal, and $v(\unit{x})$ be an extremely weak hidden variation in $b(\unit{x})$, localized with in an unknown region $\tilde{R}$ on the sphere, such that the total signal, called the observation, is given by $f(\unit{x}) = b(\unit{x}) + v(\unit{x}),\,\, \norm{v}_{\untsph} \ll \norm{b}_{\untsph}$. We assume that there are $N$ different instances~(realizations) of such a localized variation, giving us an ensemble of observations as
\begin{align}
f^j(\unit{x}) = b(\unit{x}) + v^j(\unit{x}), \qquad j = 1, 2, \ldots, N.
\end{align}
The problem under consideration is to statistically identify the presence of such anomalies (localized variations) in the source signal.

\subsection{Framework}
We compute the spatial-Slepian coefficients of the observation using the well-optimally concentrated Slepian functions with in a region $R$ on the sphere. From the linearity of the spatial-Slepian transform, we can write the spatial-Slepian transform of the $j^{\mathrm{th}}$ observation as
\begin{align}
F^j_{g_{\alpha}}(\rho) = B_{g_{\alpha}}(\rho) + V^j_{g_{\alpha}}(\rho), \qquad \alpha = 1,2,\ldots,N_R,
\label{eq:SST_f^j}
\end{align}
with statistical mean and variance given by
\begin{align}
\E\left\{ F_{g_{\alpha}}(\rho) \right\} = B_{g_{\alpha}}(\rho) + \E\left\{ V_{g_{\alpha}}(\rho) \right\},
\label{eq:mean_SST_F}
\end{align}
and
\begin{align}
\sigma^2_{ F_{g_{\alpha}}(\rho)} &= \E\left\{ \left| F_{g_{\alpha}}(\rho) - \E\left\{ F_{g_{\alpha}}(\rho) \right\} \right|^2 \right\} \nonumber \\
&= \E\left\{ \left| V_{g_{\alpha}}(\rho) \right|^2 \right\} -  \left|\E\left\{ V_{g_{\alpha}}(\rho) \right\}\right|^2 = \sigma^2_{ V_{g_{\alpha}}(\rho)},
\label{eq:var_SST_F}
\end{align}
respectively. We observe that the spatial-Slepian coefficients of the observation have the same variance as the spatial-Slepian coefficients of the localized variations, which enables us to use the sample variance across different instances, denoted by $\sv_{F_{g_{\alpha}}}$ and given by,
\begin{align}
\sv_{F_{g_{\alpha}}} = \frac{1}{N} \sum_{j=1}^N \left| F^j_{g_{\alpha}} - \frac{1}{N} \sum_{j=1}^N F^j_{g_{\alpha}} \right|^2,
\label{eq:sample_std_dev}
\end{align}
as a statistical measure for the detection of hidden localized variations in the signal at different Slepian scales $\alpha$.
\begin{figure}[!t]
	\centering
	\subfloat[$b(\unit{x})$]
	{
		\includegraphics[width=0.22\textwidth]{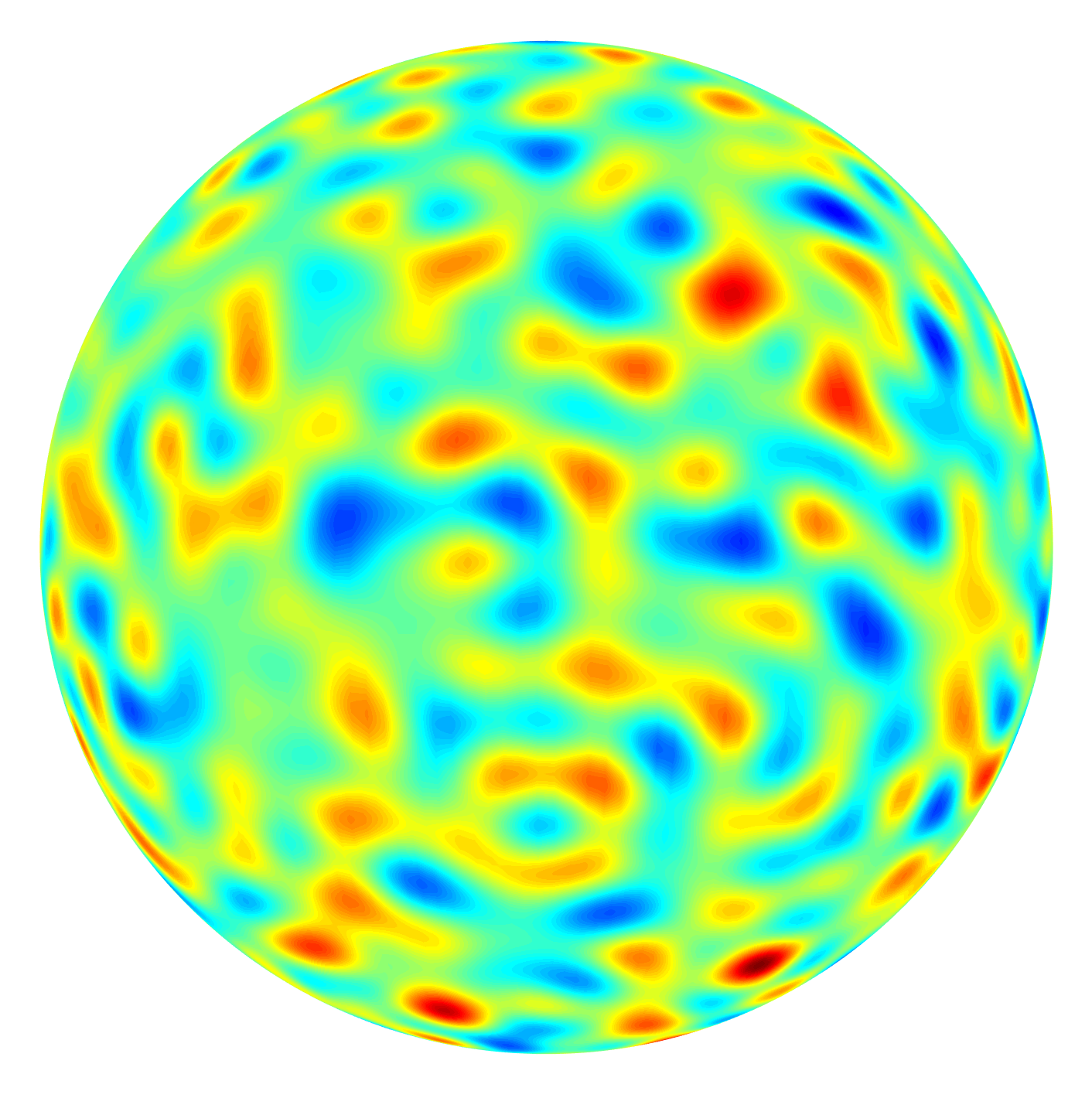}
	}\hfil
	\subfloat[$f^1(\unit{x})$]
	{
		\includegraphics[width=0.22\textwidth]{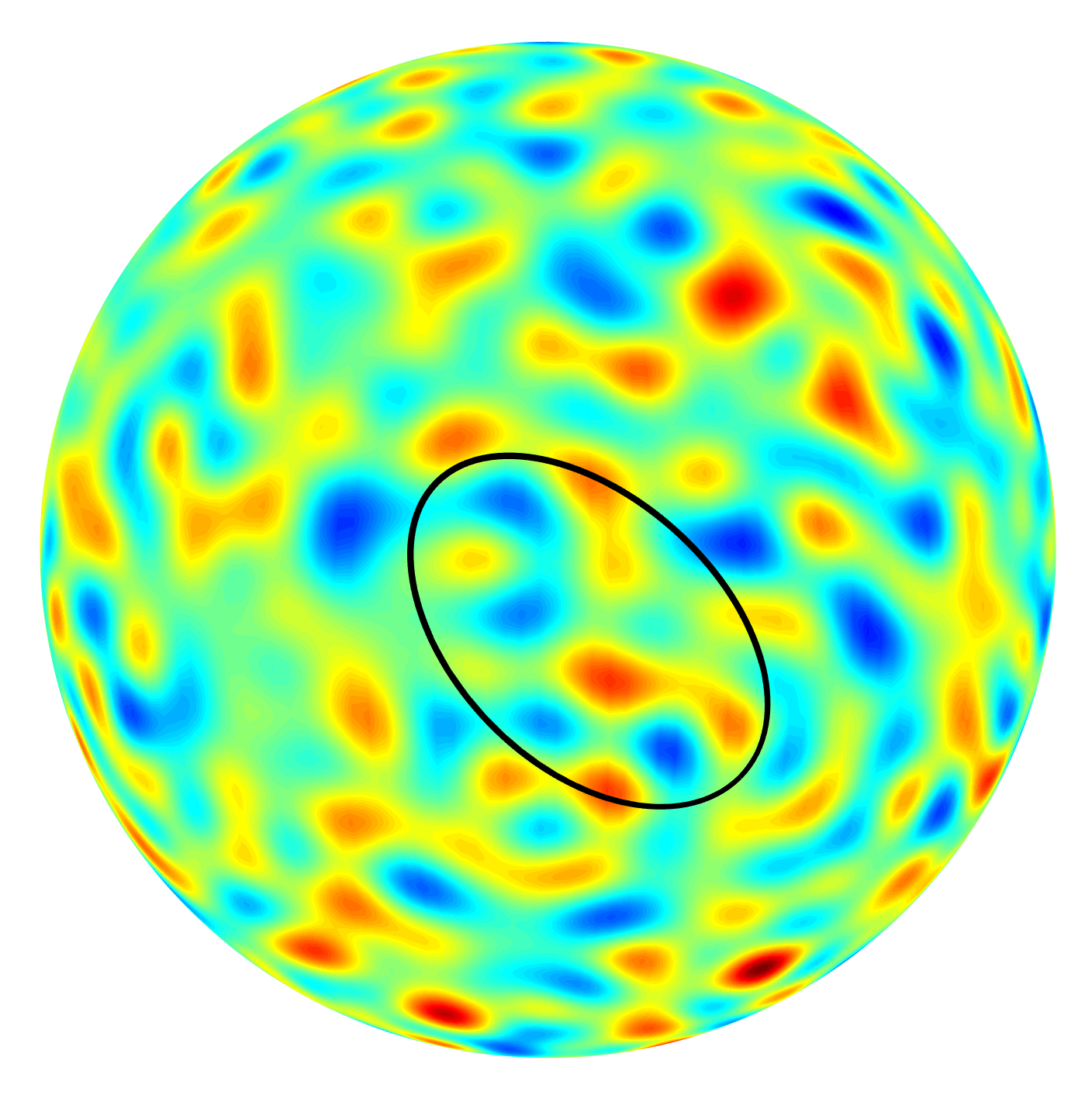}
	}\hfil	
	
	\subfloat
	{
		\includegraphics[width=0.47\textwidth]{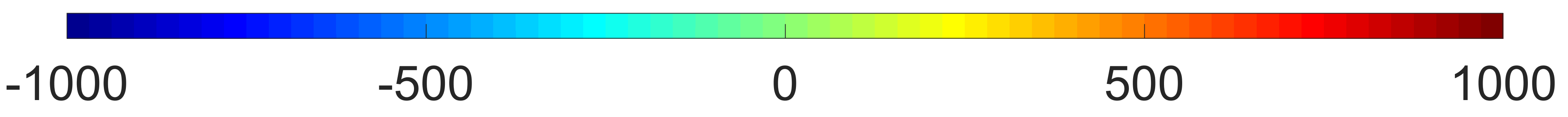}
	}\hfil
	\caption{(a) Source signal which is a realization of a zero-mean and anisotropic Gaussian process and (b) first observation that contains localized variation hidden in the source signal with in the elliptical region. Both signals are bandlimited to degree $32$.}
	\label{fig:source}
\end{figure}

\subsection{Illustration}
As an illustration, we consider a realization of the zero-mean and anisotropic Gaussian process as the background source signal $b(\unit{x})$, with bandlimit $L_b = 32$. We generate localized variations with in the region $\tilde{R}$, which is taken to be a spherical ellipse, initially aligned with $x$-axis having focus colatitude $\theta_c = 20^{\circ}$ and semi-arc-length of the semi-major axis $a = 25^{\circ}$, that is rotated by the Euler angles $\rho = (60^{\circ},90^{\circ},45^{\circ})$. The localized variations are given by
\begin{align}
v^j(\unit{x}) = \sum_{\beta = 1}^{N_{\tilde{R}}=30} a^j_{\beta} \tilde{g}_{\beta}(\unit{x}),
\end{align}
where $\tilde{g}_{\beta}(\unit{x})$ are the well-optimally concentrated Slepian functions in the region $\tilde{R}$, bandlimited to degree $L_{\tilde{g}} = 32$, $a^j_{\beta}$ are random scalars drawn from the standard normal distribution and $N_{\tilde{R}} = 30$ is the rounded spherical Shannon number for the region $\tilde{R}$. The strength of these variations is specified by the background-to-variation ratio~(BVR) defined as
\begin{align}
\bvr = 10 \log \frac{\norm{b(\unit{x})}^2}{\norm{v(\unit{x})}^2}.
\label{eq:bvr}
\end{align}
We generate $N=10$ instances of the localized variations such that BVR is $20$ dBs for each variation, yielding $N=10$ different observations on the sphere as
\begin{align}
f^j(\unit{x}) = b(\unit{x}) + \sum_{\beta = 1}^{N_{\tilde{R}}=30} a^j_{\beta} \tilde{g}_{\beta}(\unit{x}), \qquad 1 \le j \le N=10,
\end{align}
where each observation is bandlimited to degree $L_f = 32$. The source signal $b(\unit{x})$ and the observation which contains the first instance of the localized variation, i.e., $f^1(\unit{x})$, are shown in \figref{fig:source}. As can be seen, the localized variation in the highlighted elliptical region is hidden in the source signal. It must be noted that the source signal, localized variations and the spherical elliptical region $\tilde{R}$ are unbeknownst to the framework of spatial-Slepian transform.
\begin{figure*}[!t]
	\centering				
	\subfloat[$v^1(\unit{x})$]
	{
		\includegraphics[width=0.22\textwidth]{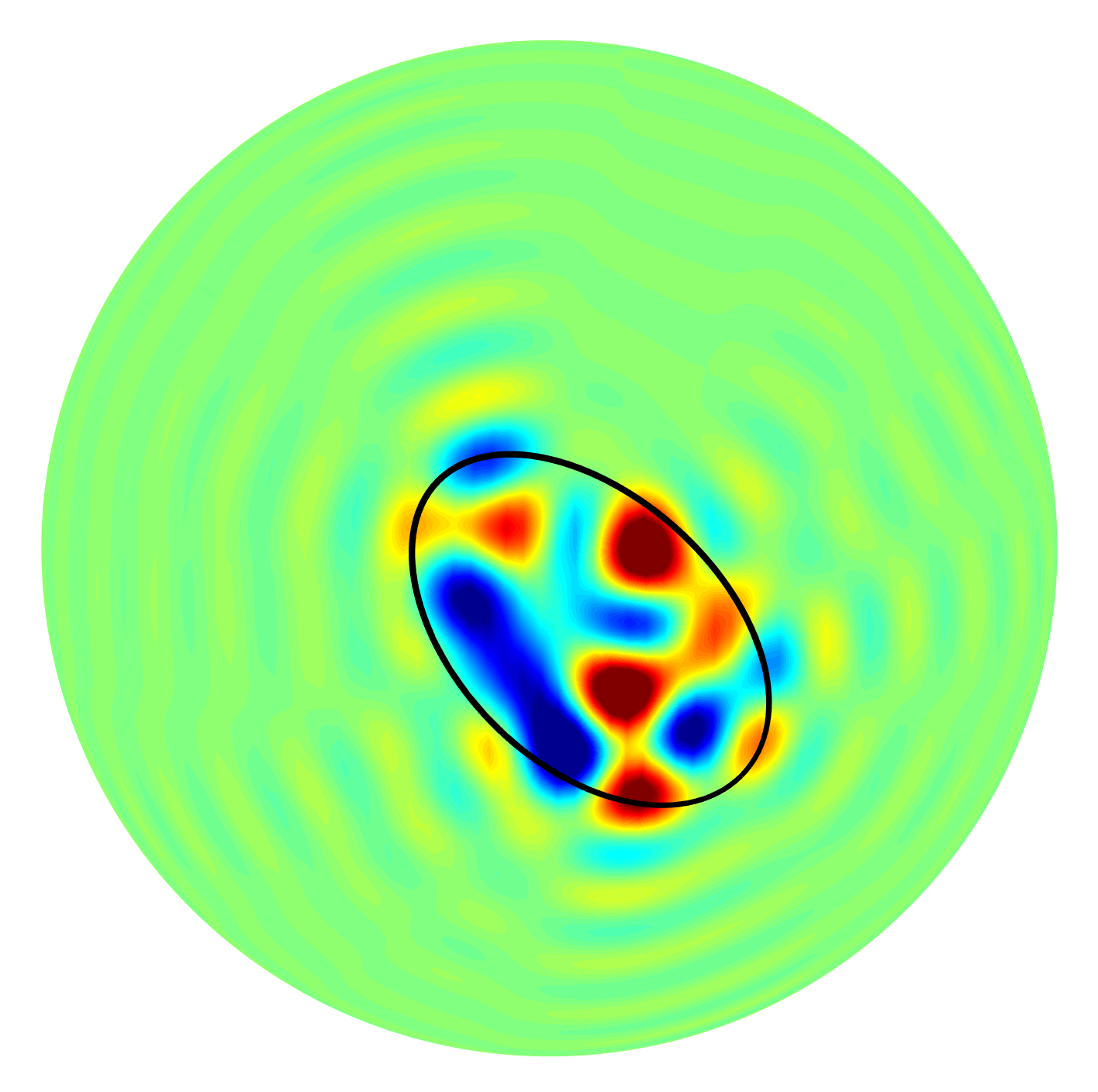}
	}\hfil
	\subfloat[$\sv_{F_{g_1}}$]
	{
		\includegraphics[width=0.22\textwidth]{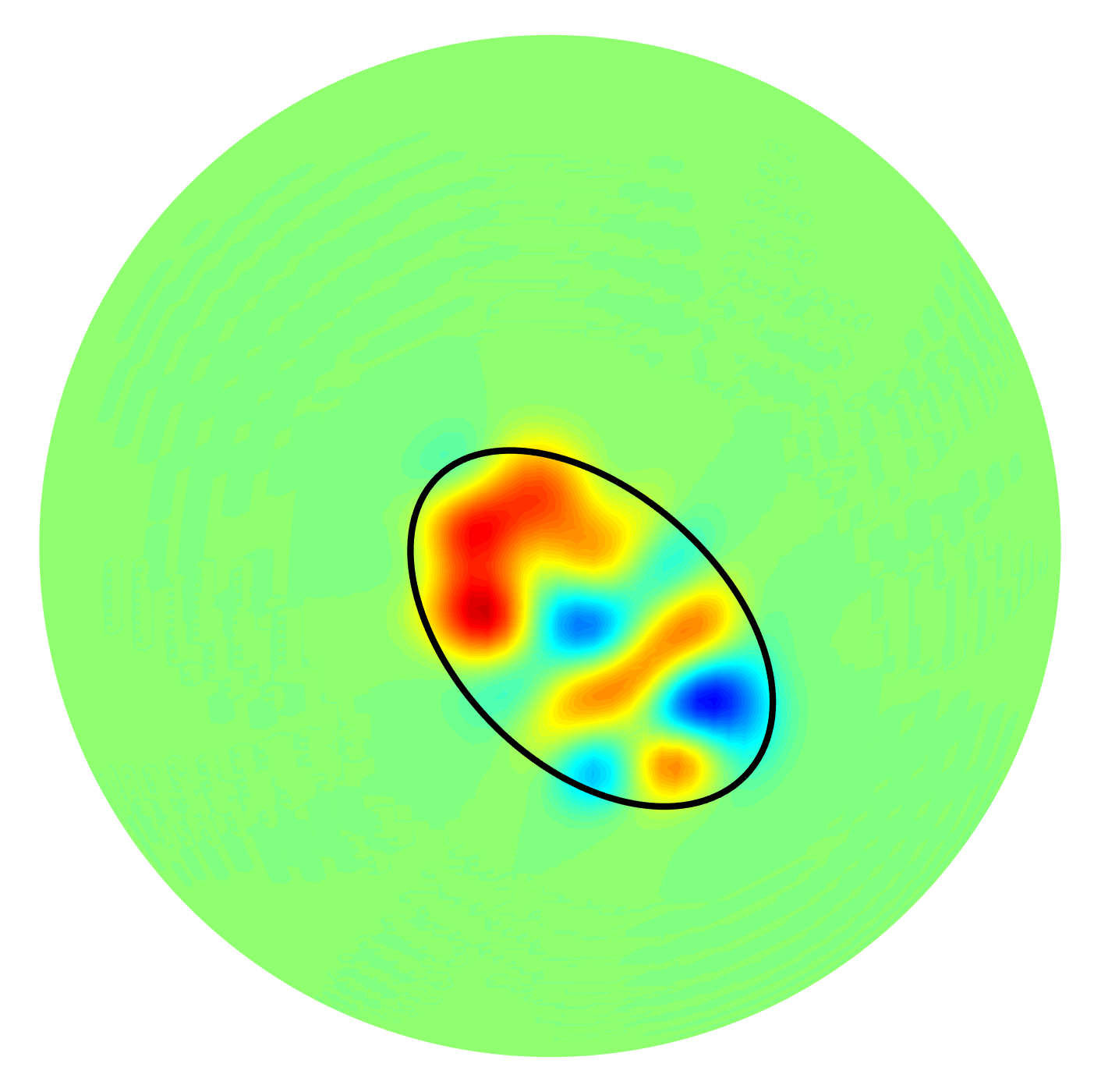}
	}\hfil
	\subfloat[$\sv_{F_{g_2}}$]
	{
		\includegraphics[width=0.22\textwidth]{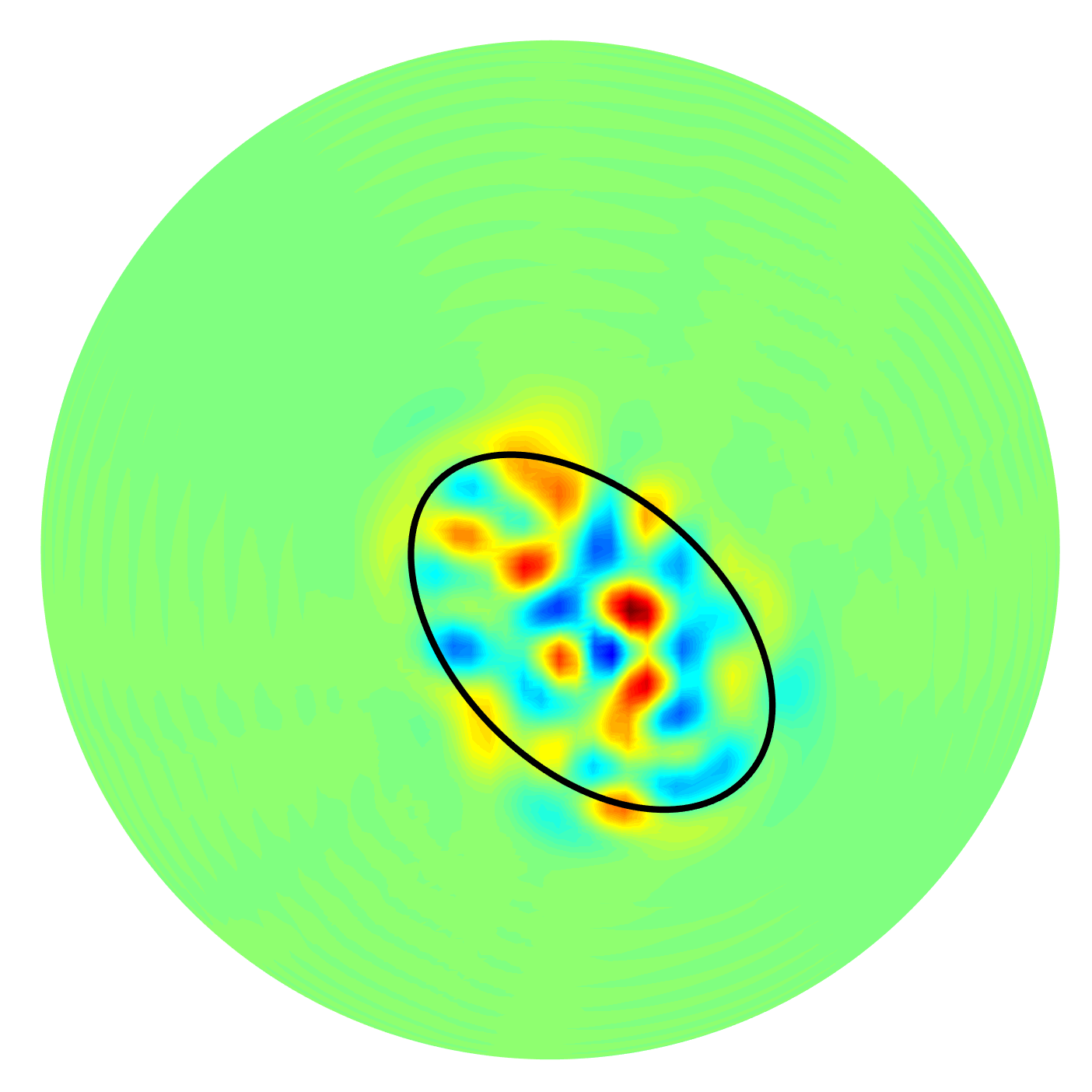}
	}\hfil
	\subfloat[$\sv_{F_{g_3}}$]
	{
		\includegraphics[width=0.22\textwidth]{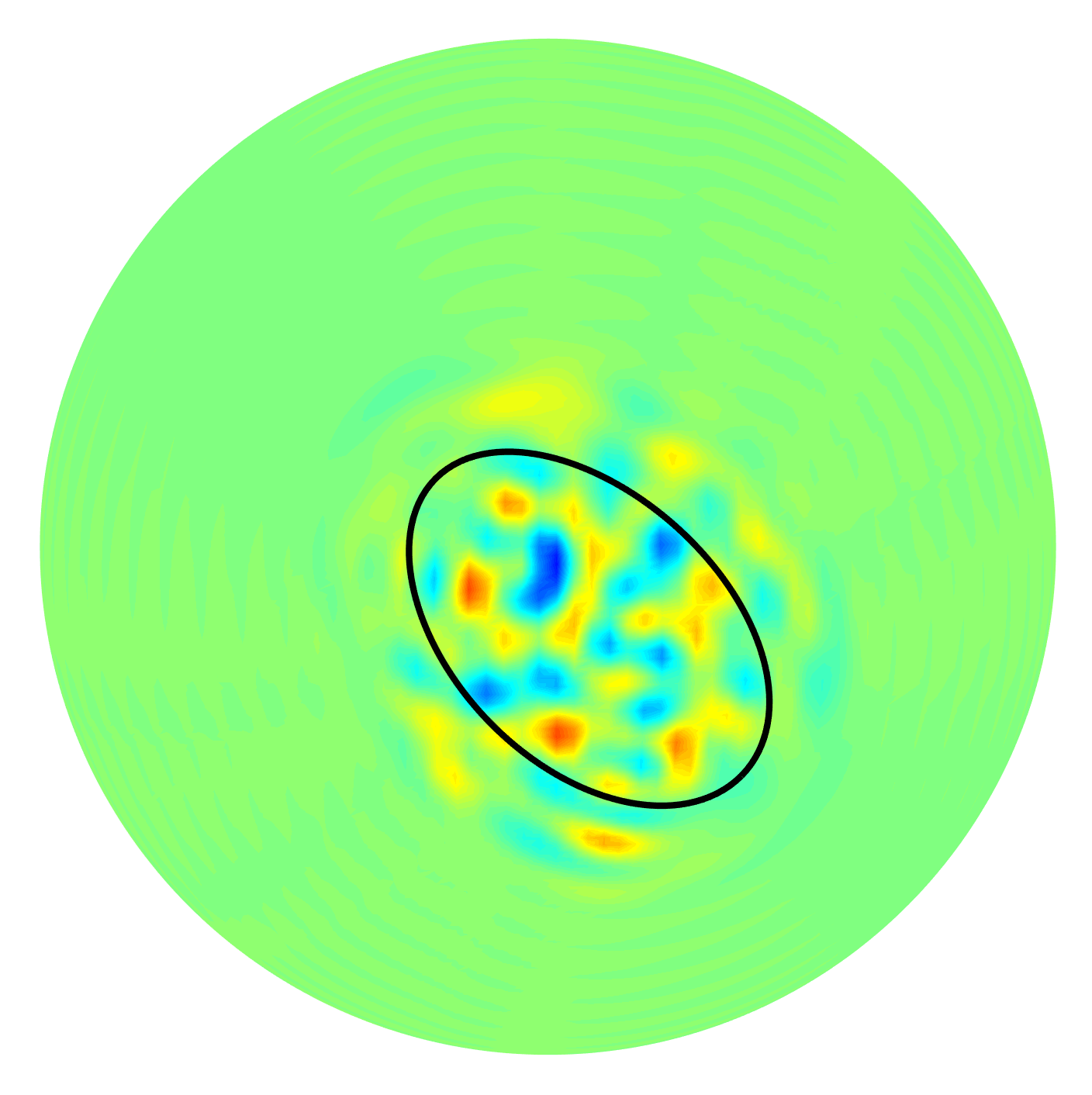}
	}\hfil
	
	\subfloat[$\sv_{\wavcoef{f}{0}}$]
	{
		\includegraphics[width=0.22\textwidth]{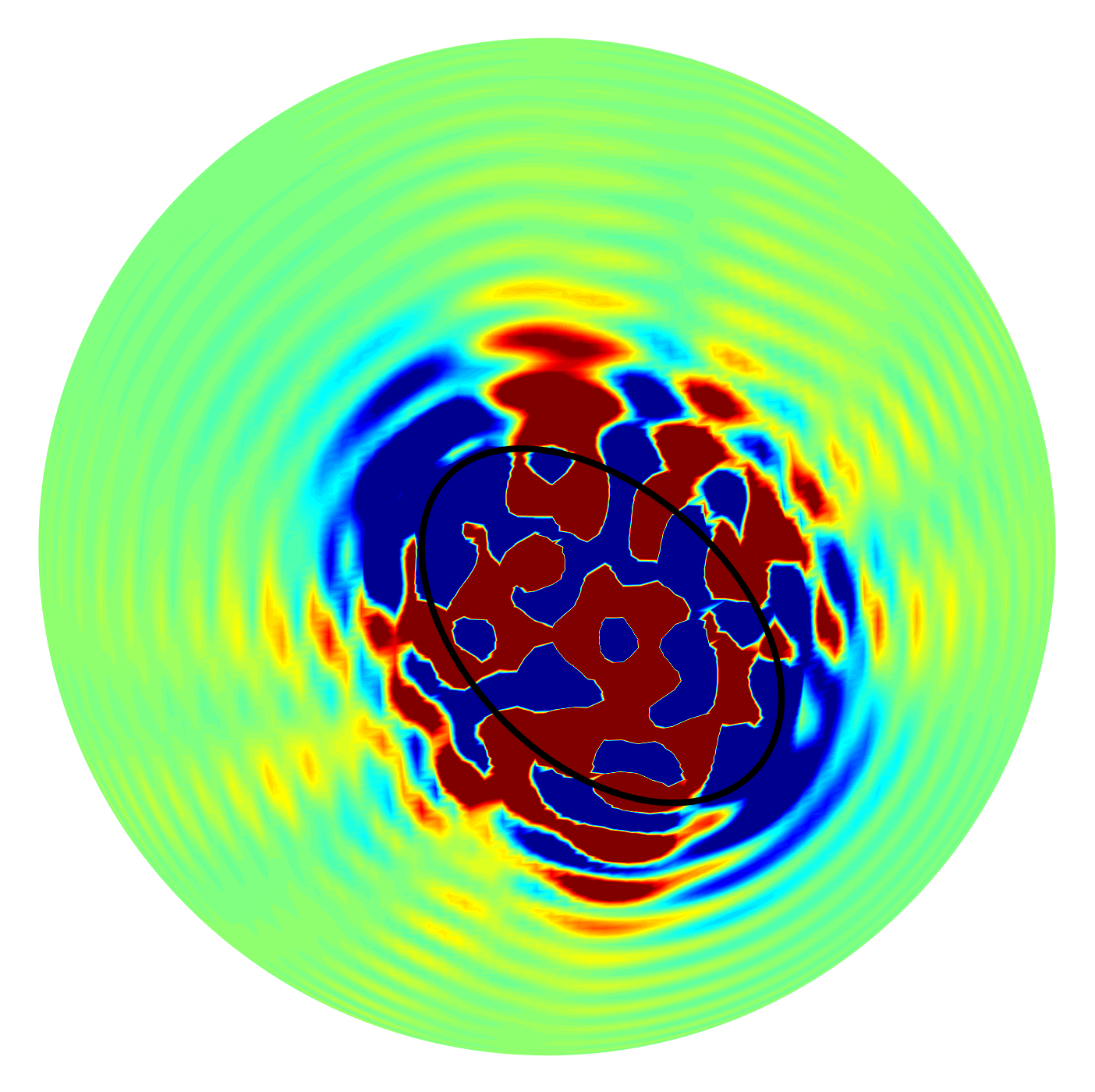}
	}\hfil
	\subfloat[$\sv_{\wavcoef{f}{1}}$]
	{
		\includegraphics[width=0.22\textwidth]{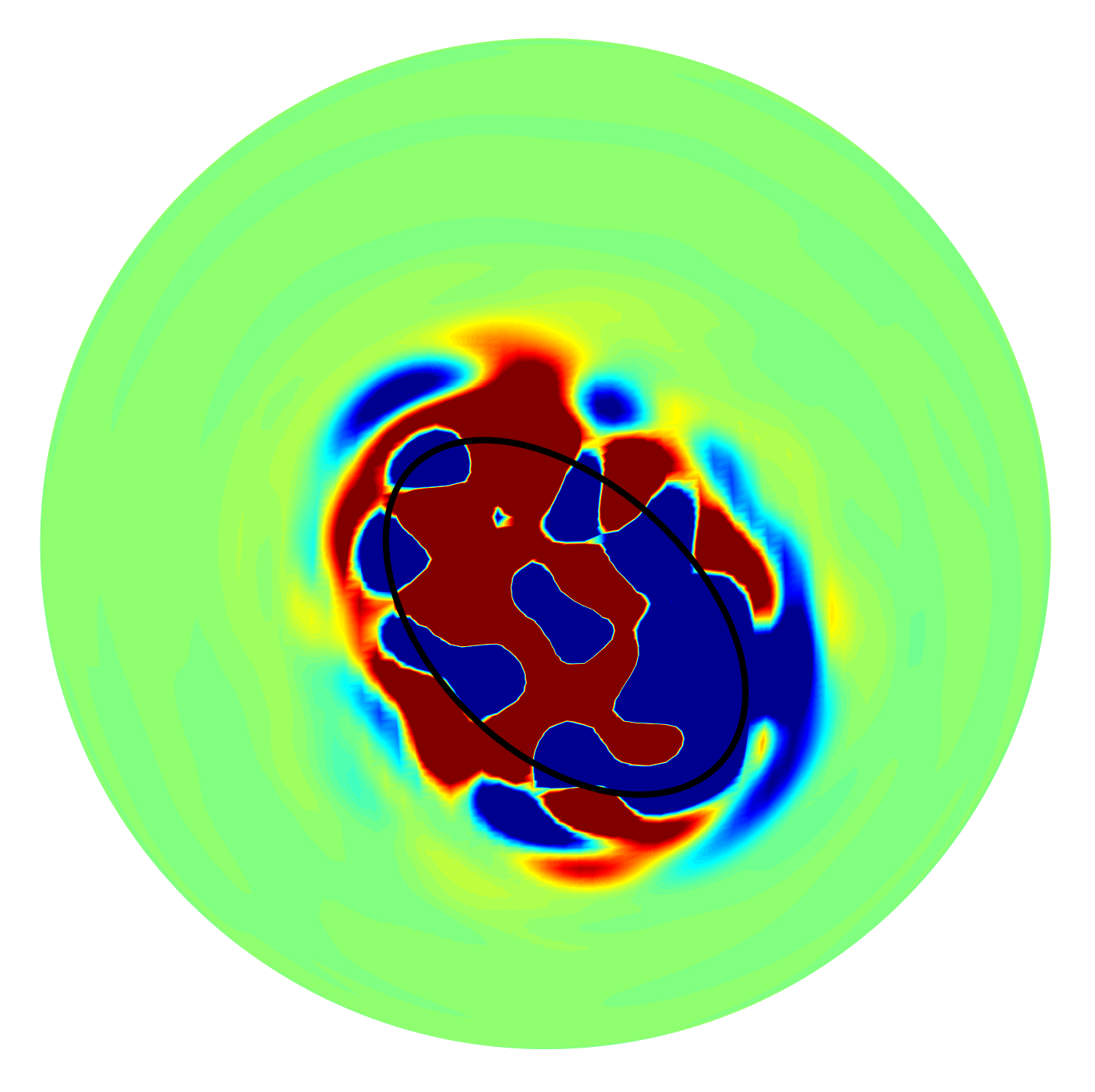}
	}\hfil
	\subfloat[$\sv_{\wavcoef{f}{2}}$]
	{
		\includegraphics[width=0.22\textwidth]{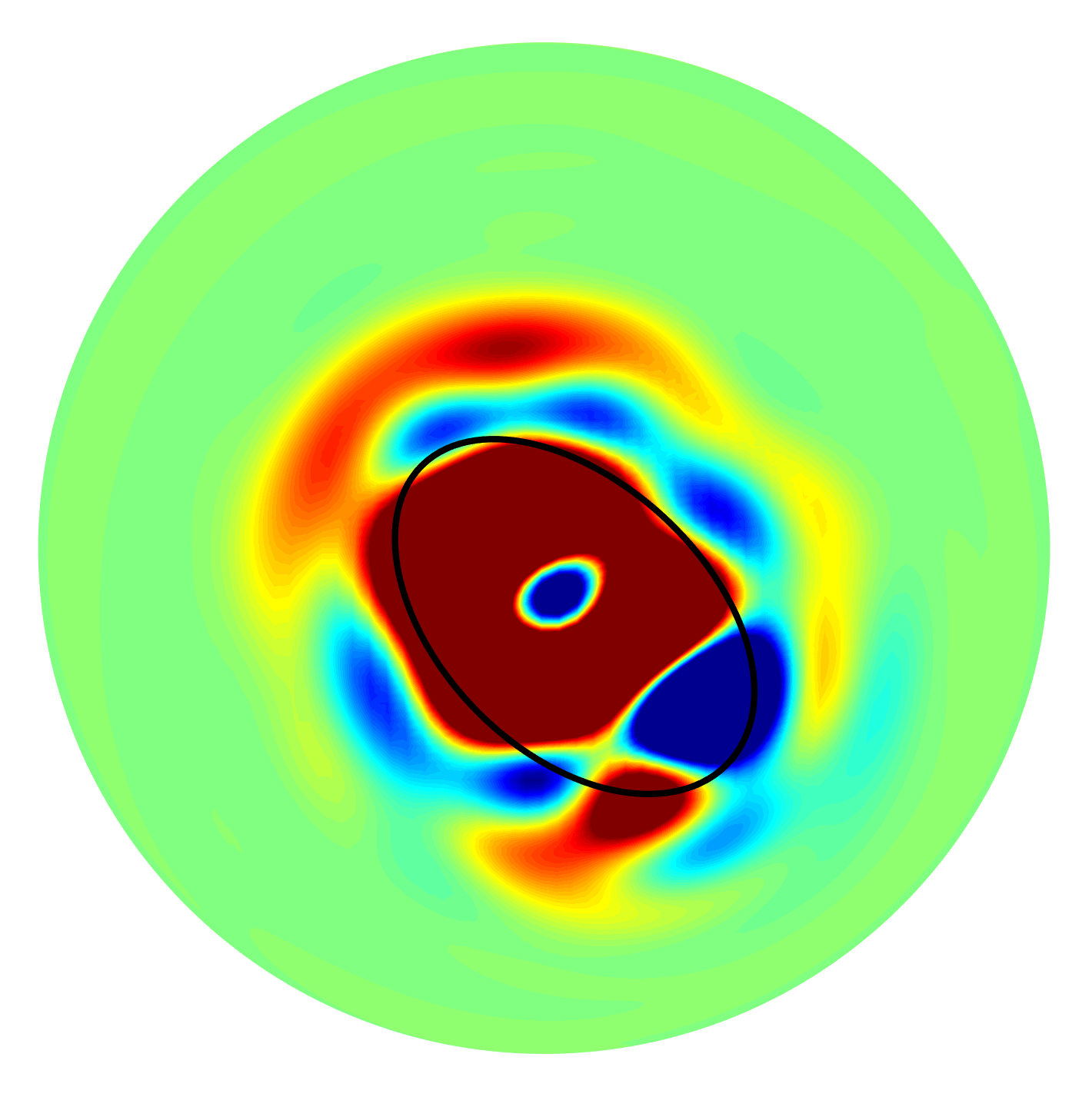}
	}\hfil
	\subfloat[$\sv_{\wavcoef{f}{3}}$]
	{
		\includegraphics[width=0.22\textwidth]{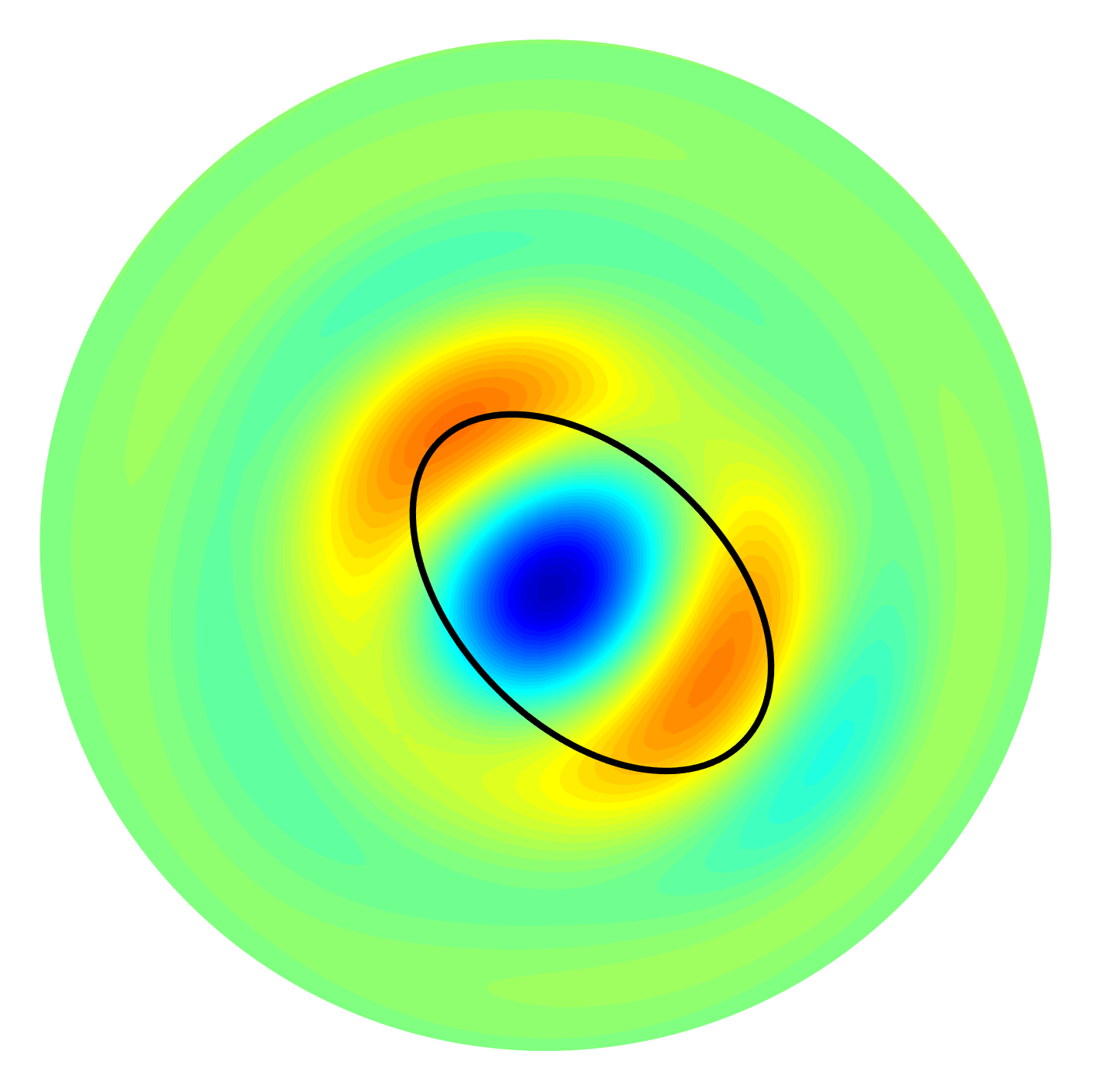}
	}\hfil
	
	\subfloat
	{
		\includegraphics[width=\textwidth]{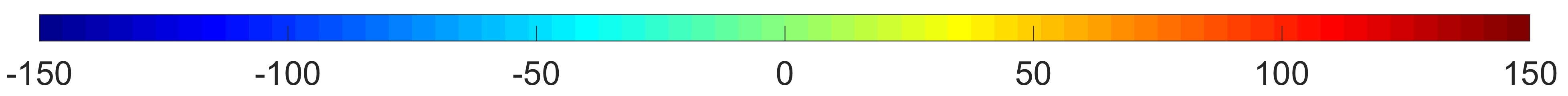}
	}\hfil
	\caption{(a) First instance of the localized variation, (b)--(d) sample variance of the spatial-Slepian coefficients, $\sv_{F_{g_\alpha}}, \alpha = 1, \ldots, N_{\pr,0}=3$. (e)--(j) sample variance of the wavelet coefficients, $\sv_{\wavcoef{f}{\res}}, \res = 0,1, \ldots, 3$. As can be seen, sample variance of the spatial-Slepian coefficients quite accurately detects the region of the localized hidden variations at each Slepian scale, $\alpha = 1, \ldots, N_{\pr,0}=3$, whereas sample variance of the wavelet coefficients yields an over-estimate of the region of localized variations. Please note that the spherical elliptical region of localized variations is unbeknownst to the framework of spatial-Slepian and wavelet transforms, and is drawn from reference only.}
	\label{fig:LVA}
\end{figure*}

The hidden variations are detected by constructing the spatial-Slepian coefficients using zonal Slepian functions over the axisymmetric polar cap region, $R$, of polar cap angle $\pr = 15^{\circ}$ with bandlimit $\LSlp = 32$, and finding the sample variance across $N=10$ different instances at each Slepian scale $\alpha = 1, \ldots N_{\pr,0}=3$. The results are shown in \figref{fig:LVA} where the unknown spherical elliptical region $R'$ is drawn for reference only.
For comparison, we also plot the sample variance of the wavelet coefficients, which are computed as~\cite{McEwen:2018}
\begin{align}
\wavcoef{f}{\res}(\rho) = \innerpS{f}{\wav{\res}} = \int_{\untsph} f(\unit{x}) \overline{(\D_{\rho} \wav{\res})(\unit{x})} ds(\unit{x}),
\label{eq:wavelet_coeff}
\end{align}
where $\wav{\res} \in \lsph$ is the wavelet function at wavelet scale $\res$. The minimum wavelet scale is $0$ and the maximum wavelet scale depends on the bandlimit, which in this case, i.e., for bandlimit $L_f=32$, is $5$. However, we choose to show the sample variance for the first $4$ wavelet scales as there is negligibly small sample variance at wavelet scales $\res=4,5$. For a detailed treatment of the wavelet transform, we refer the reader to~\cite{McEwen:2018}.

As can be seen from \figref{fig:LVA}, sample variance using the spatial-Slepian transform yields a very accurate detection of the hidden localized variations. In comparison, sample variance using the wavelet transform performs poorly; yielding an over-estimate of the underlying region of the localized variations. Superior performance of the spatial-Slepian transform is due to the fact that well-optimally concentrated Slepian functions are better suited to probe signal content locally than wavelet functions. Although, wavelet functions have been shown to exhibit good spatial localization~\cite{McEwen:2018}, unlike Slepian functions, their characteristics are not defined by the shape of the underlying region on the sphere, which makes them ill-suited for localized signal analysis on the sphere.

\section{Conclusions}
\label{sec:conc}
We have proposed spatial-Slepian transform~(SST) for the representation of a spherical signal in the joint spatial-Slepian domain, and for localized analysis of signals on the sphere. The proposed transform is similar in spirit to the wavelet transform, however, instead of using wavelet functions which cannot be adapted to a given region on the sphere, it uses bandlimited and spatially well-optimally~(energy) concentrated Slepian functions. Proposed SST probes local content of the signal, which is a direct consequence of the use of well-optimally concentrated Slepian basis functions. We have derived the constraints under which SST is invertible and have shown that well-optimally concentrated rotated Slepian functions form a tight frame on the sphere. We have also presented an algorithm for the fast computation of spatial-Slepian transform and have carried out computational complexity analysis. As an illustration, we have applied the proposed transform to the Earth topography map using the bandlimited zonal Slepian functions which are well-optimally concentrated with in an axisymmetric polar cap region on the sphere. To demonstrate utility of the proposed transform, we have also devised a framework to carry out localized variation analysis for the detection of hidden localized variations in the signal. We consider the use of proposed transform for carrying out localized signal analysis and optimal filtering as subjects of future work.

\bibliography{sht_bib}

\begin{thebibliography}{10}
\providecommand{\url}[1]{#1}
\csname url@samestyle\endcsname
\providecommand{\newblock}{\relax}
\providecommand{\bibinfo}[2]{#2}
\providecommand{\BIBentrySTDinterwordspacing}{\spaceskip=0pt\relax}
\providecommand{\BIBentryALTinterwordstretchfactor}{4}
\providecommand{\BIBentryALTinterwordspacing}{\spaceskip=\fontdimen2\font plus
\BIBentryALTinterwordstretchfactor\fontdimen3\font minus
  \fontdimen4\font\relax}
\providecommand{\BIBforeignlanguage}[2]{{%
\expandafter\ifx\csname l@#1\endcsname\relax
\typeout{** WARNING: IEEEtran.bst: No hyphenation pattern has been}%
\typeout{** loaded for the language `#1'. Using the pattern for}%
\typeout{** the default language instead.}%
\else
\language=\csname l@#1\endcsname
\fi
#2}}
\providecommand{\BIBdecl}{\relax}
\BIBdecl

\bibitem{nadeem2016spherical}
S.~Nadeem, Z.~Su, W.~Zeng, A.~Kaufman, and X.~Gu, ``Spherical parameterization
  balancing angle and area distortions,'' \emph{IEEE Transactions on
  Visualization and Computer Graphics}, vol.~23, no.~6, pp. 1663--1676, 2016.

\bibitem{michailovich:2009}
O.~Michailovich and Y.~Rathi, ``On approximation of orientation distributions
  by means of spherical ridgelets,'' \emph{{IEEE} Trans. Image Process.},
  vol.~19, no.~2, pp. 461--477, 2009.

\bibitem{Bates:2016}
A.~P. Bates, Z.~Khalid, and R.~A. Kennedy, ``An optimal dimensionality sampling
  scheme on the sphere with accurate and efficient spherical harmonic transform
  for diffusion mri,'' \emph{{IEEE} Signal Process. Lett.}, vol.~23, no.~1, pp.
  15--19, Jan. 2016.

\bibitem{bergamasco20183d}
L.~C.~C. Bergamasco, C.~E. Rochitte, and F.~L. Nunes, ``3d medical objects
  processing and retrieval using spherical harmonics: a case study with
  congestive heart failure mri exams,'' in \emph{Proceedings of the 33rd Annual
  ACM Symposium on Applied Computing}.\hskip 1em plus 0.5em minus 0.4em\relax
  ACM, 2018, pp. 22--29.

\bibitem{Bates:2015}
A.~P. Bates, Z.~Khalid, and R.~A. Kennedy, ``Novel sampling scheme on the
  sphere for head-related transfer function measurements,'' \emph{{IEEE/ACM}
  Trans. Audio, Speech, Language Process.}, vol.~23, no.~6, pp. 1068--1081,
  Jun. 2015.

\bibitem{Liu:2019}
H.~{Liu}, Y.~{Fang}, and Q.~{Huang}, ``Efficient representation of head-related
  transfer functions with combination of spherical harmonics and spherical
  wavelets,'' \emph{IEEE Access}, vol.~7, pp. 78\,214--78\,222, Jun. 2019.

\bibitem{Hoogenboom:2004}
T.~Hoogenboom, S.~Smrekar, F.~Anderson, and G.~Houseman, ``Admittance survey of
  type 1 coronae on venus,'' \emph{J. Geophys. Res.}, vol. 109, no.~2, pp.
  1--19, Mar. 2004.

\bibitem{Audet:2014}
P.~{Audet}, ``{Toward mapping the effective elastic thickness of planetary
  lithospheres from a spherical wavelet analysis of gravity and topography},''
  \emph{Physics of the Earth and Planetary Interiors}, vol. 226, pp. 48--82,
  Jan. 2014.

\bibitem{Khaki:2018}
M.~Khaki, E.~Forootan, M.~Kuhn, J.~Awange, L.~Longuevergne, and Y.~Wada,
  ``Efficient basin scale filtering of grace satellite products,'' \emph{Remote
  Sensing of Environment}, vol. 204, pp. 76--93, 2018.

\bibitem{Galanti:2019}
E.~Galanti, Y.~Kaspi, F.~J. Simons, D.~Durante, M.~Parisi, and S.~J. Bolton,
  ``Determining the depth of jupiter's great red spot with juno: A {S}lepian
  approach,'' \emph{The Astrophysical Journal Letters}, vol. 874, no.~2, Apr.
  2019.

\bibitem{Hippel:2019}
M.~v. Hippel and C.~Harig, ``Long-term and inter-annual mass changes in the
  iceland ice cap determined from grace gravity using {S}lepian functions,''
  \emph{Cryospheric Sciences, Frontiers in Earth Science}, vol.~7, no. 171,
  Jul. 2019.

\bibitem{Wieczorek:2005}
M.~A. Wieczorek and F.~J. Simons, ``Localized spectral analysis on the
  sphere,'' \emph{Geophys. J. Int.}, vol. 162, no.~3, pp. 655--675, Sep. 2005.

\bibitem{Simons:2006Polar}
F.~J. Simons and F.~Dahlen, ``Spherical {S}lepian functions and the polar gap
  in geodesy,'' \emph{Geophys. J. Int.}, vol. 166, no.~3, pp. 1039--1061, 2006.

\bibitem{Dahlen:2008}
F.~A. Dahlen and F.~J. Simons, ``Spectral estimation on a sphere in geophysics
  and cosmology,'' \emph{Geophys. J. Int.}, vol. 174, pp. 774--807, Sep. 2008.

\bibitem{Marinucci:2008}
D.~Marinucci, D.~Pietrobon, A.~Balbi, P.~Baldi, P.~Cabella, G.~Kerkyacharian,
  P.~Natoli, D.~Picard, and N.~Vittorio, ``Spherical needlets for cosmic
  microwave background data analysis,'' \emph{Mon. Not. R. Astron. Soc.}, vol.
  383, no.~2, pp. 539--545, 2008.

\bibitem{McEwen:2012}
J.~D. McEwen, S.~M. Feeney, M.~C. Johnson, and H.~V. Peiris, ``Optimal filters
  for detecting cosmic bubble collisions,'' \emph{Phys. Rev. D.}, vol.~85,
  no.~10, pp. 103--502, 2012.

\bibitem{grinter2018}
R.~Grinter and G.~A. Jones, ``Interpreting angular momentum transfer between
  electromagnetic multipoles using vector spherical harmonics,'' \emph{Optics
  letters}, vol.~43, no.~3, pp. 367--370, 2018.

\bibitem{Kennedy:2013-ICSPCS}
R.~A. Kennedy, Z.~Khalid, and Y.~F. Alem, ``Spatial correlation from multipath
  with 3d power distributions having rotational symmetry,'' in \emph{Proc. IEEE
  Int. Conf. Sig. Process. and Comm. Sys., ICSPCS}, Carrara, VIC, Australia,
  Dec 2013.

\bibitem{Alem:2015}
Y.~F. Alem, Z.~Khalid, and R.~A. Kennedy, ``3d spatial fading correlation for
  uniform angle of arrival distribution,'' \emph{{IEEE} Commun. Lett.},
  vol.~19, no.~6, pp. 1073--1076, Jun. 2015.

\bibitem{bashar:2016}
F.~Bashar, T.~D. Abhayapala, and S.~A. Salehin, ``Dimensionality of
  spatio-temporal broadband signals observed over finite spatial and temporal
  windows,'' \emph{IEEE Transactions on Wireless Communications}, vol.~15,
  no.~10, pp. 6758--6770, Oct. 2016.

\bibitem{talashila:2019}
R.~Talashila and H.~Ramachandran, ``Determination of far fields of wire
  antennas on a pec sphere using spherical harmonic expansion,'' \emph{IEEE
  Antennas and Wireless Propagation Letters}, vol.~18, no.~4, pp. 646--650,
  2019.

\bibitem{Mallat:1989}
S.~G. Mallat, ``A theory for multiresolution signal decomposition: the wavelet
  representation,'' \emph{{IEEE} Trans. Pattern Anal. Mach. Intell.}, vol.~11,
  no.~7, pp. 674--693, Jul. 1989.

\bibitem{Daubechies:1990}
I.~Daubechies, ``The wavelet transform, time-frequency localization and signal
  analysis,'' \emph{{IEEE} Trans. Inf. Theory}, vol.~36, no.~5, pp. 961--1005,
  Sep. 1990.

\bibitem{Mallat-book:2009}
S.~G. Mallat, \emph{A Wavelet Tour of Signal Processing}, 3rd~ed.\hskip 1em
  plus 0.5em minus 0.4em\relax Massachusetts, USA: Academic Press, 2009.

\bibitem{Narcowich:1996}
F.~J. Narcowich and J.~D. Ward, ``Non-stationary wavelets on the m-sphere for
  scattered data,'' \emph{Appl. Comput. Harm. Anal.}, vol.~3, pp. 324--336,
  1996.

\bibitem{Freeden:1997}
F.~Freeden and U.~Windheuser, ``Combined spherical harmonic and wavelet
  expansion – a future concept in the earth’s gravitational
  determination,'' \emph{Appl. Comput. Harm. Anal.}, vol.~4, pp. 1--37, 1997.

\bibitem{Antoine:1999}
J.-P. Antoine and P.~Vandergheynst, ``Wavelets on the 2-sphere: A
  group-theoretical approach,'' \emph{Appl. Comput. Harm. Anal.}, vol.~7,
  no.~3, pp. 262--291, 1999.

\bibitem{Starck:2006}
J.-L. Starck, Y.~Moudden, P.~Abrial, and M.~Nguyen, ``Wavelets, ridgelets and
  curvelets on the sphere,'' \emph{Astron.\ \& Astrophys.}, vol. 446, pp.
  1191--1204, Feb. 2006.

\bibitem{Wiaux:2008}
Y.~Wiaux, J.~D. McEwen, P.~Vandergheynst, and O.~Blanc, ``Exact reconstruction
  with directional wavelets on the sphere,'' \emph{Mon. Not. R. Astron. Soc.},
  vol. 388, no.~2, pp. 770--788, 2008.

\bibitem{McEwen:2018}
J.~D. McEwen, C.~Durastanti, and Y.~Wiaux, ``Localisation of directional
  scale-discretised wavelets on the sphere,'' \emph{Appl. Comput. Harm. Anal.},
  vol.~44, no.~1, pp. 59--88, Jan 2018.

\bibitem{Albertella:1999}
A.~Albertella, F.~Sans{\`o}, and N.~Sneeuw, ``Band-limited functions on a
  bounded spherical domain: the {S}lepian problem on the sphere,'' \emph{J.
  Geodesy}, vol.~73, no.~9, pp. 436--447, Jun. 1999.

\bibitem{Simons:2006}
F.~J. Simons, F.~A. Dahlen, and M.~A. Wieczorek, ``Spatiospectral concentration
  on a sphere,'' \emph{SIAM Rev.}, vol.~48, no.~3, pp. 504--536, 2006.

\bibitem{Bates2:2017}
A.~P. Bates, Z.~Khalid, and R.~A. Kennedy, ``Efficient computation of {S}lepian
  functions for arbitrary regions on the sphere,'' \emph{{IEEE} Trans. Signal
  Process.}, vol.~65, no.~16, pp. 4379--4393, Aug. 2017.

\bibitem{Kennedy-book:2013}
R.~A. Kennedy and P.~Sadeghi, \emph{Hilbert Space Methods in Signal
  Processing}.\hskip 1em plus 0.5em minus 0.4em\relax Cambridge, UK: Cambridge
  University Press, Mar. 2013.

\bibitem{Sakurai:1994}
J.~J. Sakurai, \emph{Modern Quantum Mechanics}, 2nd~ed.\hskip 1em plus 0.5em
  minus 0.4em\relax Reading, MA: Addison Wesley Publishing Company Inc., 1994.

\bibitem{Slepian:1960}
D.~Slepian and H.~O. Pollak, ``Prolate spheroidal wave functions, {F}ourier
  analysis and uncertainity-$\textsc{I}$,'' \emph{Bell Syst. Tech. J.},
  vol.~40, pp. 43--63, Jan. 1961.

\bibitem{Slepian:1964}
D.~Slepian, ``Prolate spheroidal wave functions, {F}ourier analysis and
  uncertainty -- iv: Extensions to many dimensions; generalized prolate
  spheroidal functions,'' \emph{Bell Syst. Tech. J.}, vol.~40, pp. 3009--3057,
  Nov. 1964.

\bibitem{Simons:2011}
F.~J. Simons and D.~V. Wang, ``Spatiospectral concentration in the cartesian
  plane,'' \emph{Intern. J. Geo-math.}, vol.~2, pp. 1--36, 2011.

\bibitem{Bates:2017}
A.~P. Bates, Z.~Khalid, and R.~A. Kennedy, ``{S}lepian spatial-spectral
  concentration problem on the sphere: Analytical formulation for limited
  colatitude-longitude spatial region,'' \emph{{IEEE} Trans. Signal Process.},
  vol.~65, no.~6, pp. 1527--1537, Mar. 2017.

\bibitem{Khalid:2013DSLSHT}
Z.~Khalid, R.~A. Kennedy, S.~Durrani, P.~Sadeghi, Y.~Wiaux, and J.~D. McEwen,
  ``Fast directional spatially localized spherical harmonic transform,''
  \emph{{IEEE} Trans. Signal Process.}, vol.~61, no.~9, pp. 2192--2203, 2013.

\bibitem{Trapani:2006}
S.~Trapani and J.~Navaza, ``Calculation of spherical harmonics and {W}igner
  {\it d}~functions by {FFT}. {A}pplications to fast rotational matching in
  molecular replacement and implementation into {AM}o{R}e,'' \emph{Acta Cryst.
  A}, vol.~62, no.~4, pp. 262--269, Jul. 2006.

\bibitem{Risbo:1996}
T.~Risbo, ``{F}ourier transform summation of {L}egendre series and
  {D}-functions,'' \emph{J. Geodesy}, vol.~70, pp. 383--396, 1996.

\end{thebibliography}

\end{document}